\documentclass[twocolumn, 9pt]{article}

\usepackage[english]{babel}

\usepackage[a4paper,top=1cm,bottom=2cm,left=1.5cm,right=1.5cm,marginparwidth=1.75cm]{geometry}

\usepackage{amsmath}
\usepackage{amssymb}
\usepackage{graphicx}
\usepackage[colorlinks=true, allcolors=blue]{hyperref}

\graphicspath{ {figures/} }
\usepackage{authblk} % author affiliations

\usepackage{xr}
\usepackage[table]{xcolor}
\usepackage{makecell}

\usepackage{booktabs}

\usepackage{listings}

\definecolor{codegreen}{rgb}{0,0.6,0}
\definecolor{codegray}{rgb}{0.5,0.5,0.5}
\definecolor{codepurple}{rgb}{0.58,0,0.82}
\definecolor{backcolour}{rgb}{0.975,0.95,1.0}

\lstdefinestyle{mystyle}{
    backgroundcolor=\color{backcolour},
    commentstyle=\color{codegreen},
    keywordstyle=\color{magenta},
    numberstyle=\tiny\color{codegray},
    stringstyle=\color{codepurple},
    basicstyle=\fontsize{7}{8}\ttfamily,
    breakatwhitespace=false,
    breaklines=true,
    captionpos=b,
    keepspaces=true,
    showspaces=false,
    showstringspaces=false,
    showtabs=false,
    tabsize=4,
    numbers=none,
    numbersep=5pt,
}

\title{Global continuation as a complement to traditional continuation and bifurcation analysis}
\author[1, *]{George Datseris}
\author[2, 3]{Andreas Morr}
\author[4]{Muhammed Fadera}
\author[3, 5]{J\"urgen Kurths}
\affil[1]{Department of Mathematics and Statistics, University of Exeter}
\affil[2]{Department of Mathematics, School of Computation, Information and Technology, Technical University of Munich, Boltzmannstraße 3, Garching, Germany}
\affil[3]{Potsdam Institute for Climate Impact Research, 14412, Potsdam, Germany}
\affil[4]{Global Systems Institute, University of Exeter}
\affil[5]{Department of Physics, Humboldt University, 10099, Berlin, Germany}
\affil[*]{Correspondence: g.datseris@exeter.ac.uk}

\begin{document}
\maketitle

\begin{abstract}
Multistable dynamical systems are ever-prevalent, used to model for example ecosystems, power grids,  climate elements, neurons, and more.
When perturbed, such systems may ``tip'' from one state of operation to another, often with abrupt, irreversible, and high-impact consequences in each context.
Traditionally, these systems are analysed via bifurcation diagrams, the result of a process we refer to as \emph{local continuation}, as it only captures the linear (local) system response to infinitesimal perturbations.
Local continuation requires substantial expertise, constant interventions, and may yield inaccurate assessment of the system's response to large perturbations that is crucial for tipping analysis.
To address some inherent challenges of local continuation and to provide fundamentally new  information during a continuation, this paper introduces \emph{global continuation} as a complement suitable for the study of multistability, critical transitions and real-world-oriented applications.
Global continuation finds and continues in parallel (practically) all system attractors and their response to finite perturbations by synthesising information from the whole state space, while placing a focus on the qualities or observables of a dynamical system that the practitioner cares about in context.
% Global continuation is effortless to use and troubleshoot, and unlike local continuation, it does not require deep expertise nor repeated interventions.
Global continuation does not require deep expertise and is effortless to use and troubleshoot, making it attractive to applied scientists from different disciplines.
We highlight several unique advantages that allow global continuation to complement the status quo and exemplify them through a plethora of representative examples.
Global continuation is also implemented as open source software in DynamicalSystems.jl, enhancing its accessibility.

\end{abstract}

\section{Introduction}

Multistable dynamical systems are used in science and technology to model a wide range of natural systems, ranging from the Earth's climate \cite{ArmstrongMcKay2022TippingPoints} to cell biology \cite{Zhu2022}.
Whenever natural systems admit alternative coexisting stable states or exhibit \emph{critical transitions} (also called regime shifts or tipping points \cite{Ashwin2012tipTypes}), multistable dynamical systems provide a framework for understanding and analysing the respective system behaviour.
This modelling approach has enjoyed success not only in climate science (popularized in \cite{Lenton2008} but already discussed in \cite{Stommel1961CircModel, Sellers1969, Budyko1969}) and ecology \cite{May1977, Barnosky2012},  but also in power grids \cite{Ren2015CSDElectricity}, fluid flows \cite{Mukund2018}, atmospheric flows~\cite{Herbert2020}, lasers~\cite{Rosen2010, Bhm2016}, and many more~\cite{Pisarchik2022}.
In general, critical transitions are more than bifurcations; they are fundamental changes in system behaviour that may be triggered by bifurcations or by perturbations in system parameters or state, either close or even far away from a bifurcation point (known as noise \cite{Ashwin2012tipTypes}, rate \cite{Ritchie2023RateInducedTipping}, or shock tipping \cite{Halekotte2020MinimalFatalShocks}).
% A comprehensive analysis of critical transitions in any given system therefore requires a circumspect set of techniques.

The most commonly employed approach to study dynamical models and their critical transitions is \emph{local continuation}, hitherto called just ``continuation''.
Continuation is a numerical technique for tracking a system’s steady state(s) and other properties as parameter(s) are varied.
Local continuation, specifically, achieves this by tracking an individual fixed point or limit cycle over primarily a single parameter and recording the local (linearized) stability of the state \cite{Dankowicz2013}.
The key principles behind local continuation were developed largely in the 1970s and 1980s, leading to the implementation of the established AUTO software \cite{Auto2007, Krauskopf2007NumericalContinuationMethods}.
Nevertheless, this single-state analysis requires intervention and repetition in the case of multistability, as the practitioner is blind to dynamics away from the attractor. Also, the focus on linearized stability is often inadequate for capturing the response to realistic (finite-sized) perturbations relevant to studying critical transitions \cite{Menck2013BasinStability}.

In this paper, we propose that a novel technique we name \emph{global continuation}, is emerging as a complement to traditional analyses of dynamical systems.
Global continuation automatically finds and tracks (practically) all attractors of a dynamical system, as well as their stability well beyond the linear regime, while ignoring unstable states. The inherent focus on multistability and stability beyond linearity makes the method particularly suitable for the study of critical transitions~\cite{Morr2026}.
A key accessibility feature of global continuation is its simplicity: it does not require deep expertise beyond basic familiarity with the concept of attractors, and it is straightforward to use and troubleshoot.
% Its accessibility is further enhanced by its software implementation in DynamicalSystems.jl as we will expose in the Methods.
Global continuation is focused on pragmatic applications; its output revolves around the qualities or observables of a dynamical system that the practitioner cares about,
aligning naturally with real-world phenomena that the dynamical system attempts to capture.
%, and the response of the system to finite perturbations. Both of these align naturally with real-world phenomena the dynamical system attempts to capture, while mathematical details such as how to label a particular bifurcation are left for local continuation to provide if need be.

In \S\ref{sec:global_cont} we introduce the core philosophy behind global continuation.
We start with a high-level exposition of the differences between local and global continuation \S\ref{sec:global_cont_highlevel}, formalise the method in \S\ref{sec:global_cont_formal}, and highlight its unique features in more detail in \S\ref{sec:global_cont_unique}.
Then in \S\ref{sec:examples} we provide some key example applications that highlight the usefulness of global continuation in diverse scenarios.
A plethora of additional examples are provided in the Supplementary Information.
We conclude in \S\ref{sec:discussion} with a discussion of the limitations of global continuation and its usefulness for the wider scientific community.

\section{Global continuation}
\label{sec:global_cont}

\subsection{Local and global continuation}
\label{sec:global_cont_highlevel}

\begin{figure*}
    \centering
    \includegraphics[width=1\linewidth]{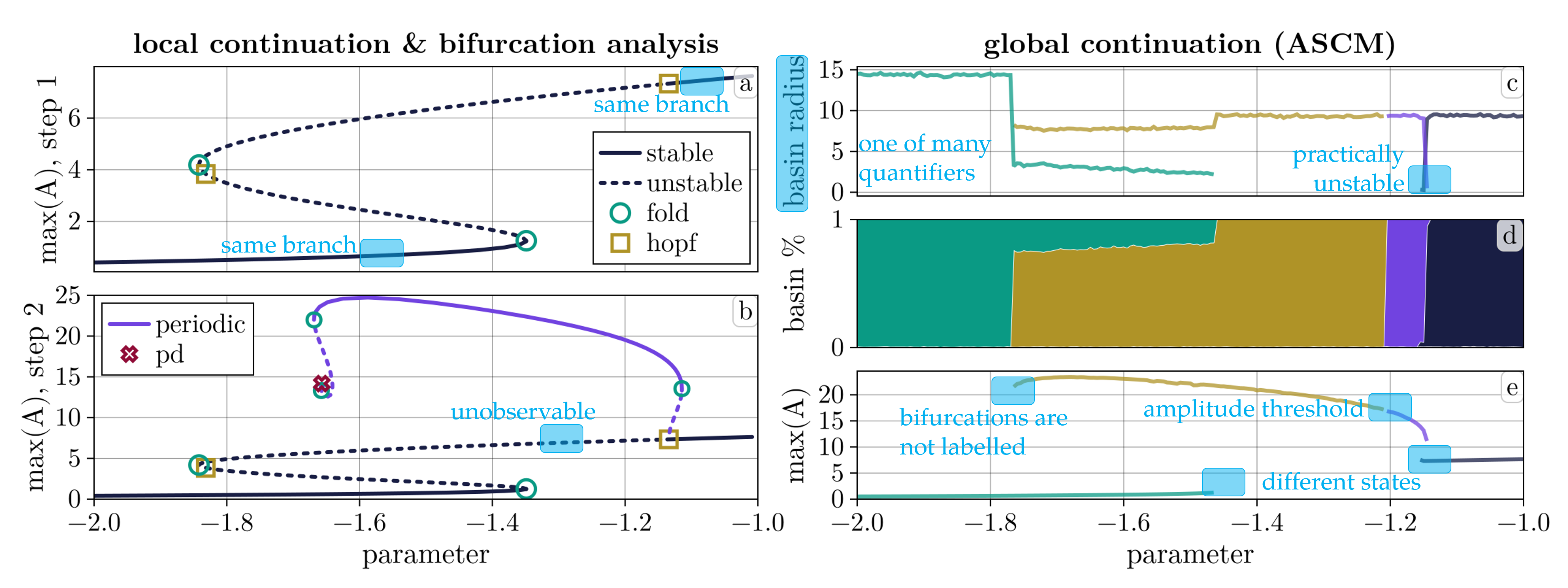}
    \caption{\textbf{High level comparison between traditional (local) continuation and bifurcation analysis and global continuation.} Both techniques are applied to a three-dimensional neuronal mass model (details provided in Supplementary Information, \ref{sec:model_details}). Panels (a, b, e) plot the maximum of one of the model variables. Panel (b) should be compared with panel (e), the latter showing only the attracting states of the system. Light blue colour is used for annotations.
    }
    \label{fig:comparison}
\end{figure*}

Local and global continuation are both techniques for tracking properties of a dynamical system as parameter(s) change.
What properties to track, however, and how to do so, is fundamentally different between the two techniques.
To showcase this, we apply both in Fig.~\ref{fig:comparison} to an exemplary neuronal model, with key differences annotated in light blue.
An extensive comparison between the two methods is provided in \S\ref{sec:comparison}.

The output of local continuation is ``branches'': curves of fixed points or limit cycles in the joint state-parameter space. Branches are found primarily by using variants of Newton's method~\cite{Dankowicz2013}, which transforms the original dynamical system and therefore does not preserve the system's flow.
Each branch is tracked individually and regardless of whether a state is stable or unstable.
For tracking a new branch, the practitioner has to manually intervene, to facilitate at a minimum the starting point and the algorithm to use (as different algorithms must be used for fixed points or limit cycles).
Along a branch, system stability is recorded via the linear response (Jacobian eigenvalues) of the system.
This output is mathematically relevant, but physically speaking there are some challenges: 1) different stable states, that may even be co-existing, may be classified in the same branch, even if their physical context may be completely different (e.g., a snowball or warm-climate planet \cite{Budyko1969}); 2) the reporting of unstable states complicates subsequent analysis, however unstable states are not observable in the real world, at least without some invasive or feedback control~\cite{Sieber2008, Wallner2026}, nor are they related directly to basin boundaries in non-trivial high dimensional systems;
3) the linear response is often an inadequate characterization of realistic system stability~\cite{Menck2013BasinStability}.
Therefore, after local continuation is performed, additional methodologies must be employed to study and understand multistability, basin boundaries, or transitions between attractors.

Global continuation does not provide branches, nor does it operate on the joint state-parameter space.
It tracks over a prescribed parameter curve.
Over this parameter curve, it finds and tracks (practically) all system attractors in parallel, instead of focusing on one individually.
Finding attractors is done in a fundamentally different way than local continuation, by employing a basin map, see \S\ref{sec:global_cont_formal} and Methods \ref{met:basin_maps}.
Then the output is the different co-existing attractors, their properties, and their basins' properties, over the prescribed parameter curve.
This way, attractors that are different are presented as different, regardless of what history of unstable branches may or may not connect them.
Additionally, global continuation allows the practitioner to be explicit about when two attractors should be different via a process called ``matching'', see~\S\ref{sec:global_cont_formal} and Methods \ref{met:matching}.
In the example of Fig.~\ref{fig:comparison}, we impose that attractors that have oscillation amplitude above a prescribed threshold are fundamentally different than those with amplitude below the threshold, due to e.g., corresponding to a physically-relevant level of neuronal activity.

Besides the attractors, various other properties are provided, which can be any function of the attractors or their (sampled) basins of attraction.
For example, the basin stability~\cite{Menck2013BasinStability} (Fig.~\ref{fig:comparison}d), or the average basin radius (Fig.~\ref{fig:comparison}e), which is the average distance of all points in the basin from the attractor centroid.
If this quantity is very small, the attractor is unstable for most practical purposes.
Indeed, we can confirm that the fixed point attractor which starts at $p\approx -1$ is effectively unstable before the Hopf bifurcation point indicated in Fig.~\ref{fig:comparison}b.
A major downside of global continuation is the lack of refined identification and labelling of bifurcations: in global continuation, a bifurcation is ``detected'' whenever an attractor appears or disappears. For more details see \S\ref{sec:comparison}.

\subsection{Global continuation formalization}
\label{sec:global_cont_formal}

We assume a dynamical system described by some state vector $\mathbf{u}$. The time evolution of $\mathbf{u}$ is governed by the flow $\Phi$, a function that maps any point in the state space to its future, $\Phi^{t}(\mathbf{u(0)}; p) \to \mathbf{u}(t)$.
The details of $\Phi$ do not matter, i.e., we are not limited to a particular type of system such as continuous or discrete time, deterministic or stochastic, autonomous or not.
In all cases $\Phi$ is parameterized by some parameter(s) $p$ that do not change as the system evolves (or, for physical applications, change much slower than the timescales of the system).
Practically, $\Phi$ is approximated by numerical integration schemes, for which the practitioner can specify the accuracy up to machine precision.
From $\Phi$ we define the key structure necessary for global continuation, the \emph{basin map} $\mathcal{B}(\mathbf{u}(0)) \to j\in \mathbb{N}$.
$\mathcal{B}$ evolves an initial condition $\mathbf{u}_0$ and maps it to its corresponding basin of attraction, enumerated by integers.
Typically, basins of attraction are defined around attractors $\mathcal{A}_j$, sets that generalize fixed points and capture the long-term behaviour of a dynamical system~\cite{Datseris2022NonlinearDynamicsJulia, Ott2002}.
We do not mathematically formalize attractors in this work, because it is the construction of $\mathcal{B}$ that decides how an attractor is defined.
Some basin maps also permit mapping initial conditions to objects that are not formally attractors, but generally a class of operating or functional response, or metastable states, see Methods for details~\S\ref{met:basin_maps}.
No matter the definition however, due to the computational nature of our framework, everything is considered in finite time.

In global continuation one uses $\mathcal{B}$ to find (practically) all unique attractors $\mathcal{A}_j$, and then continues all of them in parallel over a parameter curve, while recording several quantifiers of local and nonlocal stability.
The simplest quantifier of nonlocal stability is the \emph{basin fraction} (also known as \emph{basin stability}~\cite{Menck2013BasinStability}) $F_j$: the probability that a random initial condition will converge to $A_j$.
Beyond this, a plethora of other quantifiers of local and nonlocal stability, or any other property of the attractors and/or their basins can be tracked, as listed in Methods~\ref{met:quantifiers}.

\begin{figure*}
    \centering
    \includegraphics[width=1\linewidth]{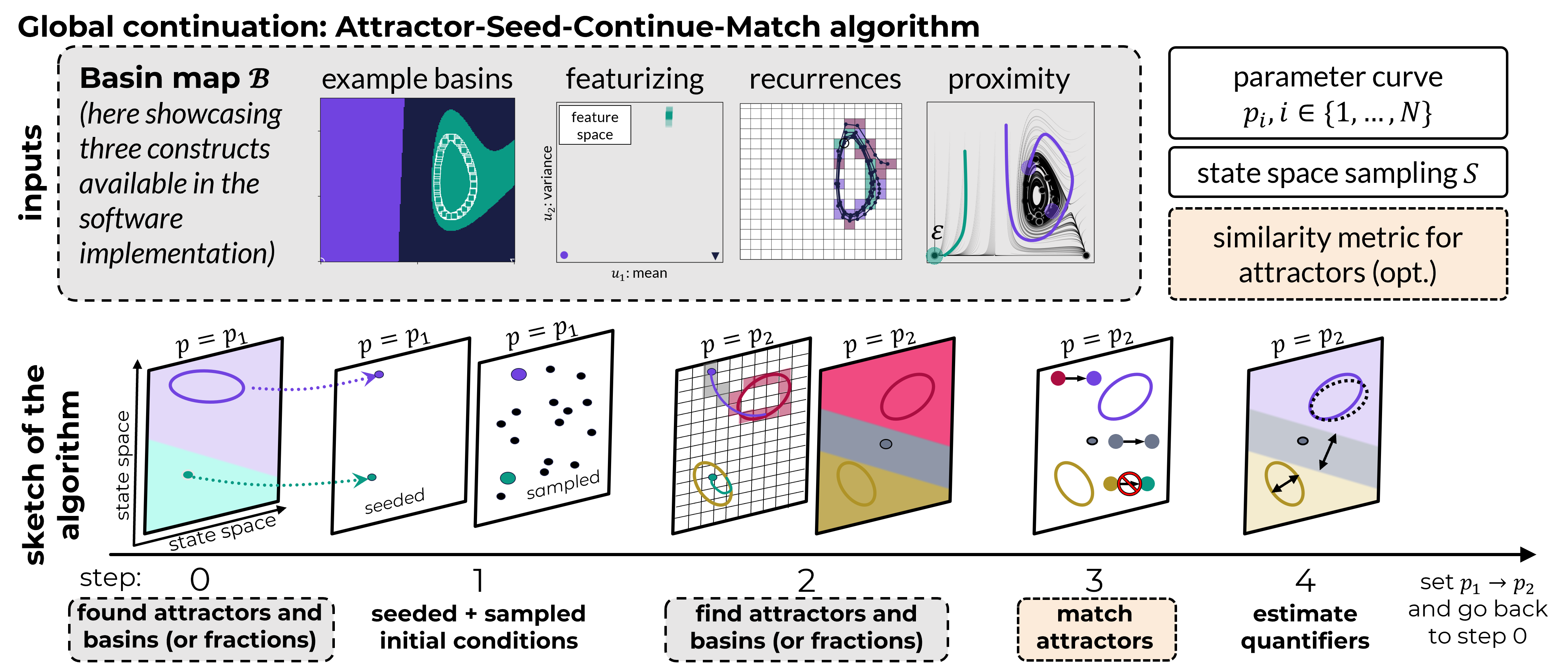}
    \caption{\textbf{Global continuation algorithm Attractors-Seed-Continue-Match (ASCM).}
    \textbf{Inputs}: (1) $\mathcal{B}$, the basin map that maps initial conditions to integers (enumerating the attractors they converged to). $\mathcal{B}$ references a dynamical system which depends on parameters $p$.
    (2) $p_i$, a discretized parameter curve existing in an arbitrary parameter space (i.e., not just a single parameter).
    (3) $S$, instructions on how to sample the state space for initial conditions (either a distribution to be sampled or a prescribed set).
    The sample points can also correspond to perturbations from a prescribed state.
    $S$ can be provided as a vector $S_i$, having different value for each parameter $p_i$.
    (4) optional input: what measure of similarity to use to decide whether an attractor at parameter $p_{i+1}$ is ``the same'' as a previous one at $p_{i}$. By default this uses the distance of attractor centroids, see Methods \S\ref{met:matching}.
    \textbf{Algorithm}: \emph{Step 0:} We start at $p_1$ by applying $\mathcal{B}$ to find attractors, the fractions of their basins of attraction, and any other quantifier desired.
    \emph{Step 1:} From each found attractor we \emph{seed} initial conditions by collecting a couple of random points on each attractor. The reason being that if an attractor continues to exist (i.e., no bifurcation occurs), initial conditions close to it will likely converge to its slightly modified form in the new parameter(s), even if the basin of attraction has become very small.
    These seeded initial conditions are then combined with many more sampled from $S$.
    \emph{Step 2:} The parameter(s) is now incremented, $p_1 \to p_2$, and all collected initial conditions are mapped to attractors using $\mathcal{B}$, obtaining for free the respective basin fractions. This step may find new attractors. The probability to find new attractors is proportional to their basin fraction, and depends on the provided $S$ (see \ref{sec:comparison}).
    \emph{Step 3:} Each attractor gets assigned an arbitrary unique integer, which leads to the next step: ``matching''. This crucial step is what establishes this method as a continuation: it ensures that attractors that are ``similar enough'' are also labelled with the same integer (colour in the plot). In contrast to local continuation, the practitioner gets to decide what ``similar enough'' means, see Methods.
    In the sketched example plot limit cycles are not matched to fixed points, irrespectively of their proximity in state space, because they have different meaning in the context of the dynamical system's operation (illustrative example).
    \emph{Step 4:} In the final step information that has already been obtained is processed to estimate any other quantifier the practitioner requests.
    In the plot we highlight the Lyapunov exponent, the attractor variance, and the average diameter of the basin of attraction (each is estimated for each attractor).
    \textbf{Outputs}: global continuation provides (1) the matched attractors at each step of the curve $p_i$, (2) their corresponding basin fractions, (3) a plethora of other quantifier(s) that can be extracted from the sampled basins or attractors, see Methods \S\ref{met:quantifiers} for what is available out-of-the-box in the software implementation.
    An example output is shown in Fig.~\ref{fig:comparison}, panels (c-e).
    }
    \label{fig:global_continuation}
\end{figure*}

In this paper, we present a specific algorithm for global continuation, which we call \emph{Attractors-Seed-Continue-Match} (ASCM), while alternative algorithms are discussed in the Supplementary Information.
The algorithm details are outlined in Fig.~\ref{fig:global_continuation}.
The explanation of the main principles behind global continuation and ASCM is best understood in parallel with the Figure, and is thus exposed in the Figure caption.

From the algorithm it becomes clear that the basin map $\mathcal{B}$ is a key structure: it decides for what dynamical systems the global continuation can be applied to.
If a basin map $\mathcal{B}$ can be defined for a system in a way that is sensible for the practitioner, global continuation and all of its offerings can then be obtained.

In the Methods section \S\ref{met:basin_maps} we discuss three versions of $\mathcal{B}$ already existing in the software implementation. One of the versions, based on the recurrences algorithm of Ref.~\cite{Datseris2022BasinsAttraction}, requires no expert knowledge of the dynamical system or prior analysis.
The practitioner only needs to provide a region in the state space, arbitrarily large, that may contain attractors.
From there further refinement and troubleshooting is straightforward.
Limitations of global continuation are discussed in \S\ref{sec:limitations}.

\subsection{Unique features of global continuation}
\label{sec:global_cont_unique}
Global continuation provides a plethora of unique features that enable fundamentally new analyses, or dramatically simplify existing ones.
Each feature is exemplified in various examples, either in the main text or in the Supplementary Information (SI).

\begin{enumerate}
    \item \textbf{Automatically, and in parallel, tracks and separates multiple co-existing attractors.} While doing so, the practitioner can optionally decide what it means for attractors to be different or not. Examples: all.
    \item \textbf{Preserves system flow and basin structure}. This makes it possible to calculate (during the continuation) basin volumes and critical perturbations that can bring system state across different basins.
    % In contrast, local continuation does not preserve the basins of attraction, see \S\ref{sec:comparison}.
    Examples: \ref{ex:rnn}, \ref{ex:hilbert_predprey}, \ref{ex:lorenz84}.
    \item \textbf{Fully supports chaotic attractors}, which are believed to be prevalent in climate and complex systems. Examples: \ref{ex:lorenz84}.
    \item \textbf{Continues several quantifiers of stability along with the attractors}. This provides a novel view into analysing resilience of dynamical systems and allows for new ways to understand and detect critical transitions. Examples: \ref{ex:rnn}, \ref{ex:hilbert_predprey}, \ref{ex:lorenz84}, \ref{ex:aggregation}.
    \item \textbf{Can be aggregated across attractors that share similar properties.} This places focus on the observable characteristics of the system rather than the dynamical system behaviour. Examples: \ref{ex:aggregation}, \ref{ex:bayesian}, \ref{ex:lossy_network}. Aggregation details are discussed in \ref{met:aggregation}.
    \item \textbf{Tracks over arbitrary prescribed parameter curves}. This allows the practitioner to conveniently focus on the way parameters change in the real-world or a real experiment, as well as incorporating parameter uncertainty via Monte Carlo simulations. Examples: \ref{ex:hilbert_predprey}, \ref{ex:clouds}, \ref{ex:bayesian}.
    \item \textbf{Identifies parameter regions satisfying requested dynamical properties.} These properties can be anything ranging from multistability, properties or existence of particular attractors, or properties of the basins. Useful in identifying regions with desired functional state or safe operating spaces. Examples: \ref{ex:hilbert_predprey}.
    \item \textbf{Basins fractions can be used for Bayesian parameter inference.} This is because the basin fraction is essentially the probability to observe a state with a particular behaviour. Example: \ref{ex:bayesian}.
    \item \textbf{Uninterrupted by any kind of local or global bifurcations.} Global continuation does not stop when encountering, e.g., complex global bifurcations, that may occur in e.g., high dimensional network models.
    Regardless of bifurcations it continues until the parameter curve is tracked in full without the need of interventions. Examples: \ref{ex:rnn}, \ref{ex:lossy_network}.
    % \item \textbf{Can track metastable states via the creation of appropriate basin maps}. Example given in SI examples \texttt{metastable}.
    %\item \todo{Not sure if will include:} \textbf{Algorithmically formalizes convergence and similarity between attractors.} This removes ambiguity and allows the creation of new algorithms. E.g., phase tipping return diagrams, or treating meta-stable states as attractors, allowing their systematic study and continuation.
    \item \textbf{Relies only on the existence of a basin map}. Thus, it is easily applicable to dynamical systems that may not fall into classes of predefined bifurcation types or be differentiable.
    This is particularly relevant for applications in engineering or machine learning, that often involve discontinuities in the vector field or its derivative.
    This also removes the need for e.g., multiple shooting algorithms to find correct initial conditions to identify limit cycles: any initial condition that converges to the cycle under $\Phi$ is a ``correct'' one.
    Examples: \ref{ex:rnn}, \ref{ex:lossy_network}, \ref{ex:engineering}.
\end{enumerate}

\section{Key examples}
\label{sec:examples}

\subsection{Opening the blackbox in recurrent neural networks}
\label{ex:rnn}
Global continuation provides a natural framework for understanding the decision-making process of trained recurrent neural networks (RNNs), which are widely used for time-series tasks and form the basis of many modern AI systems such as ChatGPT. Despite their success and ubiquity \cite{Karpathy_2015}, RNNs remain largely black-box models, particularly since it is unclear why adequately trained RNNs make errors. Recent work \cite{ceni2020interpreting, ashwin2024transitions} has shown that in sequence-to-sequence classification, the decision-making process of minimally trained RNNs, such as echo state networks, can be interpreted as input-driven transitions between fixed points, each of which may correspond to one of the classes. Some attractors may exist in the state space but may be task-irrelevant, and thus transitions to them may decrease performance, depending on how easy it is to transition to them. This is quantified by the input-driven excitability thresholds: the minimum perturbation required by inputs to make a transition from one attractor to another.

\begin{figure}
    \centering
    \includegraphics[width=\linewidth]{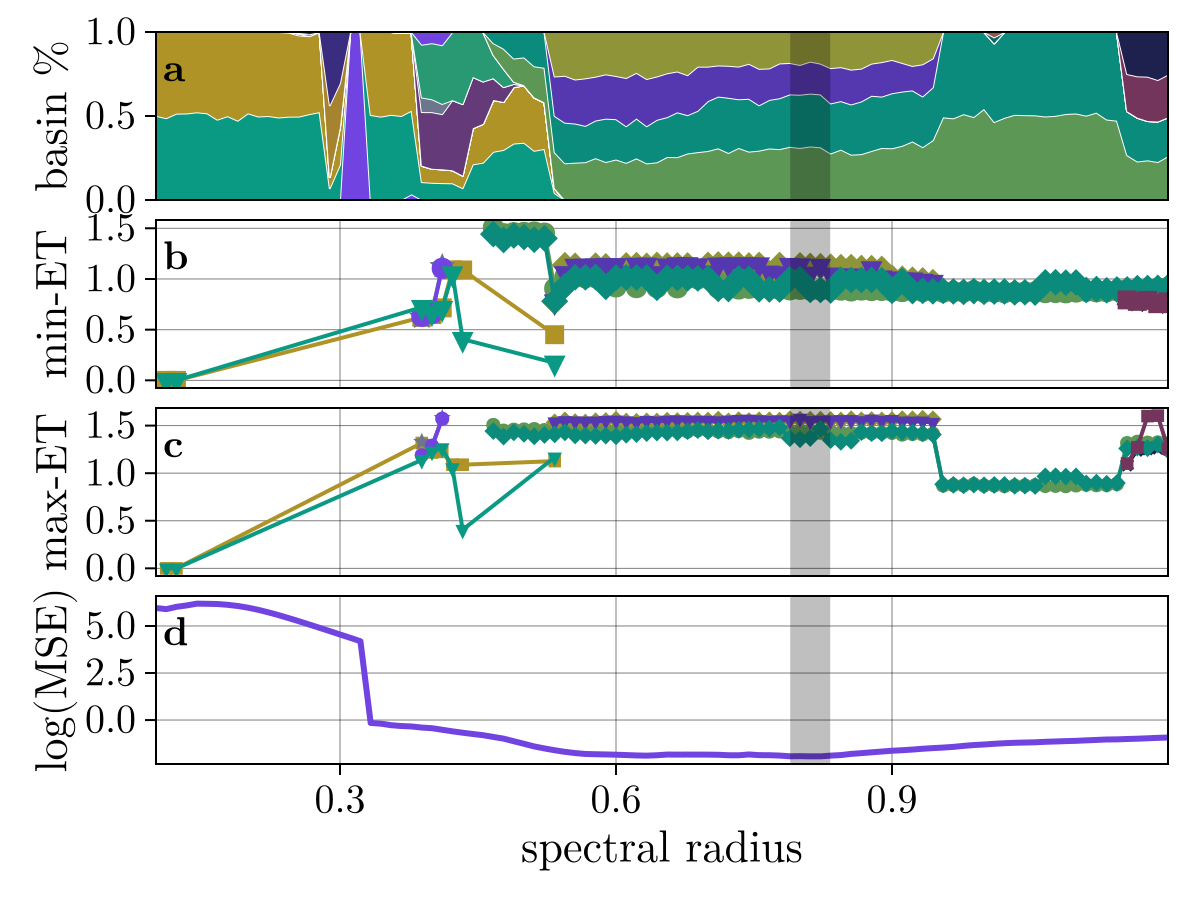}
    \caption{\textbf{Global continuation of an Echo State Network (ESN) of 500 nodes while preserving and tracking its basin structure.}
    a: basin fractions,
    b, c: minimum and maximum input-driven excitability threshold (ET) for fixed point attractors (sharing colours with a),
    d: mean squared error (MSE, measure of the ESN performance).
    A shaded black region denotes the spectral radius that minimizes MSE.
    This continuation was done using the featurize-and-group basin map (Methods, \ref{met:basin_maps}). Similar results were obtained using recurrence-based map.
    }
    \label{fig:highlight_rnn}
\end{figure}

Here we use global continuation to continue excitability thresholds of fixed points of echo state networks (ESN, a specific variant of RNNs discussed more in the supplement, \ref{sec:model_details}). We study the response versus the spectral radius, which is the primary hyperparameter for ESNs, and governs the stability of long-term nonautonomous dynamics. The ESNs were trained on the $2$-bit flip-flop task, a four-class sequence-to-sequence classification problem with piecewise-constant inputs \cite{jaeger2001echo}. The $2$-bit flip-flop task is to independently memorise, for each component of a bivariate time series taking values in $[-1, 0, 1]$, when it was last non-zero.
Figure~\ref{fig:highlight_rnn} shows the output of the global continuation along with the (minimized) trained error of the ESN.

Global continuation helps us understand exactly why the ESN error is minimised where it is, from a dynamical systems perspective.
It reveals that error is minimized when the following conditions all coincide: (1) the state space is populated only by fixed point attractors whose number coincides with the number of classes the ESN needs to predict; (2) the input-driven excitability thresholds are not too small or too large; (3) the basin fractions corresponding to the fixed points are all sufficiently large and roughly of equal size.
Wherever these conditions are not satisfied, ESN performance is suboptimal (error is not minimal).

For example, when the spectral radius is less than $0.4$, the state space is populated by a plethora of other attractors, some being periodic.
For spectral radius approaching 1, both the number of attractors is incompatible with the classes, but also their excitability thresholds are lower (i.e., their stability decreases).
Even when the number of fixed-point attractors returned to the desirable number, with a spectral radius of $1.1-1.2$, some fixed points may allow transitions to others under very small perturbations (panel c, dark red) and to others under relatively large perturbations (panel d, dark red).
The former case may result in ``wrong" transitions, while the latter may result in under exploration of fixed points corresponding to other classes. Both of these result in a decrease in performance as seen in panel d.

Using local continuation to perform such analysis is impractical.
Firstly, for each value of the spectral radius, the ESN needs to be retrained (with details about training in \ref{sec:model_details}). Thus, backtracking during a local continuation (e.g., tracking an unstable branch over already explored parameter values) is very expensive.
% It also comes at no immediate benefit, as the unstable fixed points cannot be used directly to infer the excitability thresholds.
The same argument applies when tracking different fixed points, again over the same parameters.
In global continuation, a pre-trained sequence of ESN matrices can be continued over once, tracking all existing stable fixed points (or other states), saving multiple orders of magnitude in computational cost.
Global continuation can also calculate and track the excitability thresholds, as it preserves the ESN's flow $\Phi$, while this must be done manually after-the-fact in local continuation.
Lastly, local continuation is also inefficient to use for small spectral radii of 0.3-0.6, where a large sequence of local and global bifurcations occurs, necessitating multiple manual interventions.
More details on continuation applications on ESNs can be found in~\cite{fadera2025attractors}.

\subsection{Identifying parameter subspaces yielding desired system properties}
\label{sec:feature_parameter_subspaces}
\label{ex:hilbert_predprey}

A key feature of global continuation is its ability to be performed over arbitrary prescribed curves in a parameter space.
A curve that is of particular relevance here is a (finite) Hilbert curve.
It allows covering a high-dimensional parameter space with arbitrary density.
The global continuation over this curve can then be straightforwardly processed to partition or label parts of the parameter space according to desired properties of the attractors or basins contained within.
By continuing over a Hilbert curve in multiparameter spaces, matching is still performed as usual, thus providing a unique identity of alternative attractors not over only one, but over multiple dimensions.

\begin{figure*}
    \centering
    \includegraphics[width=1\linewidth]{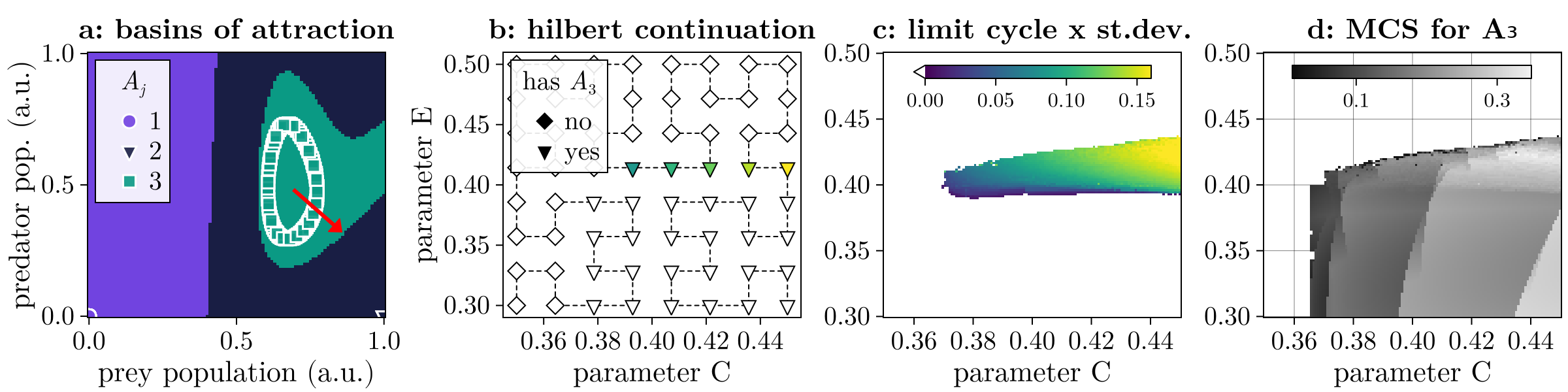}
    \caption{\textbf{Multiparameter global continuation and parameter region identification/segmentation.} Hilbert curve based global continuation of a simple predator prey model with two parameters $C, E$. See \S\ref{met:code} for a code snippet exactly producing this continuation. a: basins of attraction for $C, E = (0.4, 0.4)$, with the attractors $A_j$ over-plotted.
    b: Global continuation of the model over a two-dimensional parameter space, covered efficiently using a Hilbert curve. The Hilbert curve is plotted with a dashed line. At each point on the curve, a marker demonstrates whether a coexistence attractor exists or not.
    The colour of the marker indicates the limit cycle amplitude, if any.
    c: Limit cycle amplitude, if any, now plotted as a heatmap using a higher density Hilbert curve. Any conceivable property of the attractors or their basins could be plotted instead, which is demonstrated in panel d, which plots the minimal critical shock of the coexistence attractor (magnitude of red arrow in panel a).
    Panels \textbf{b, c} share the colour map.
    This continuation was done using the featurize-and-group basin map, see \ref{met:basin_maps}.}
    \label{fig:hilbert_continuation}
\end{figure*}

An exemplary showcase of this is demonstrated in Figure~\ref{fig:hilbert_continuation} for a predator-prey model.
This model, which represents the most basic form of an ecosystem, can have up to three attractors depending on the model parameters.
However, only one of these (green in Fig.~\ref{fig:hilbert_continuation}) corresponds to a healthy coexistence state where both species are alive.
It is therefore important to efficiently identify parameter regimes where such a  state exists, and if it does, whether the populations fluctuate (limit cycle) or are static (fixed point).
Using a global continuation over a Hilbert curve practically trivializes this.
After the continuation has covered the (in this scenario two-dimensional) parameter space, a simple filtering operation is performed: all parameter combinations that yield at least one attractor with both populations fluctuating with finite amplitude are kept for further analysis, as in Fig.~\ref{fig:hilbert_continuation}c.
This yields the parameter region(s) of interest.

Global continuation over multiparameter spaces is particularly useful for identifying ``safe operating regimes'' in terms of both parameter values as well as system response, while allowing the practitioner to define what ``safe'' means.
For example, in Fig.~\ref{fig:hilbert_continuation}d, the minimal critical shock for the coexistence state is estimated throughout the parameter plane.
A safe operating regime could be defined as parameters which correspond to a MCS exceeding a threshold corresponding to a sufficiently high level of ecological resilience in terms of species number perturbations (such as harvesting).

Achieving similar output with local continuation techniques is possible, but substantially more complex, while entirely missing the information of critical shocks.
In the Supplement (\ref{sec:comparison_multiparameter}), we provide a detailed comparison between local and global multiparameter continuation.

% Given a parameter configuration for a dynamical system, we can use $\mathcal{B}, \mathcal{A}$ to learn how many attractors there are, and what are their types. Global continuation can connect (match) this information across a multi-dimensional parameter space. Afterwards, this information can be processed and aggregated, so that the parameter space is divided into regions that fullfill certain properties. For example, bistable parameter regions where at least two attractors co-exist, or chaotic regions where a chaotic attractor exists, or context-relevant regions where system attractors fullfill certain statistical criteria, etc.

% Comments: local continuation only works within parameter limits: given a parameter limit, finds the fixed point(s) within the limit. For two parameters, cannot be generalized to an area. Instead, a further constraint is added: finds the bifurcation point in a 2D parameter space. But by now it is well established that system stability can change meaningfully far away from bifurcations. E.g., perturbations of less and less magnitude may cause tipping in the system before the bifurcation even occurs. This is why you want to prescribe a parameter curve to track over.

\subsection{Tracking over prescribed parameter curves while incorporating parameter uncertainty}
\label{ex:clouds}

A realistic dynamical system will likely depend on several parameters, and it is often necessary to understand how the system behaves when changing several of them.
Crucially, a range for most parameters must often be explored, due to uncertainty surrounding them in the application context.
In addition, some parameters may change in a correlated and prescribed manner, reflecting e.g., an experimental setup.
To our knowledge there is no way to meaningfully incorporate such possibilities in local continuation.
% To do this with local continuation one must employ multi-parameter variants~\cite{HENDERSON2002, Dankowicz2013}, which is substantially more complicated that typical continuation.
%, and has not seen many applications in multistable systems. This is likely due to the complexity of disentangling multistability, as local continuation does not attribute unique labels to individual attractors. But even after disentangling alternative states, incorporating uncertainty in the transition between states is impossible.
Global bifurcation is fundamentally different.
The practitioner provides a curve on an arbitrarily-high parameter space, and the continuation occurs over this curve.
This makes it trivial to focus on parameter curves (or even regimes, via the usage of Hilbert curves) that are relevant for the target application, while Monte Carlo sampling of alternative parameter curves can incorporate uncertainty.

As an example to illustrate the point and its usefulness, we highlight the work of \cite{Datseris2026Clouds}, who developed a multistable five-dimensional dynamical system representing transitions from a high cloud coverage state to a low cloud coverage state.
The study identified 6 model parameters ($U,D,CO2,\Delta_+T, RH_+, T_{FTR}$), corresponding to broader environmental conditions, that are all key in deciding whether the model will ultimately permit a high coverage cloud state.
To study the system response to climate change, and hence estimate the likelihood of a transition, multiple model parameters need to be varied either with or without correlations.
Uncertainty regarding climate change means that the values of each parameter, and the rate they vary, generates several different climate change scenarios one has to model.
For global continuation each of these scenarios is simply one prescribed parameter curve that the continuation will run over.
The continuation will automatically estimate which attractors exist, which do not, and what are their properties, without any hassle.

\begin{figure}
    \centering
    \includegraphics[width=\linewidth]{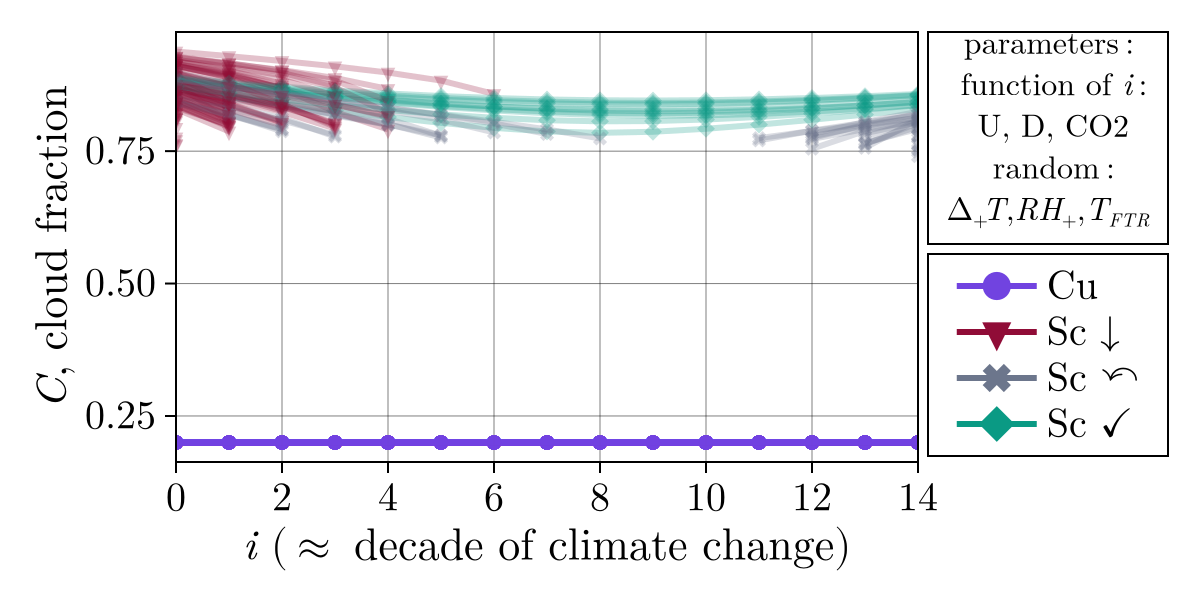}
    \caption{\textbf{Multiple continuations with prescribed parameter changes and parameter uncertainty.} Multi-parameter continuations of a cloud model representing the study of climate change scenario on cloud state transitions between high and low cloud cover (from Ref.~\cite{Datseris2026Clouds}).
    In the plot, 100 continuations are visualized with transparent colour, and each of the continuation segments is given a unique colour and marker according to the state it represents: low cloud cover (Cu), persistent high cloud cover (Sc $\checkmark$), high cloud cover loosing stability during the continuation (Sc $\downarrow$) and high cloud cover regaining stability after it has lost it (Sc $\curvearrowleft$)
    This continuation was done using the recurrences-based basin map, see \ref{met:basin_maps}.
    }
    \label{fig:cloud_model}
\end{figure}

In Figure~\ref{fig:cloud_model} we showcase some continuation results of this cloud model.
During the continuation, three parameters of the model are varied, all three changing linearly while the parameter curve $p_i$ index $i$ increases.
The figure however shows the result of 100 continuations over-plotted.
That is because the remaining three parameters have a finite degree of uncertainty. To address this, they are sampled randomly at the start of the continuation.

Counting the percentage of red curves (Sc $\downarrow$) in Fig.~\ref{fig:cloud_model} is a probabilistic estimate of a high cloud cover collapse due to climate change (for this model and simulation setup). And it is straightforward to perform this calculation, due to the simplicity of the output of global continuation.
Note that for this exact setup, one can obtain almost the same information with direct numeric integration, as the state of interest is a single fixed point.
However, this relies on being able to find this fixed point regardless of the 6 variable parameters, and also assuming the same ``bifurcation structure'' as parameters change.
Global continuation not require these assumptions.
Not only it finds the state automatically for any parameter configuration (if it exists), but can also be performed for an arbitrary complex state of interest, or even multiple states that have a similar function, by utilizing aggregation, without worry about what bifurcations may occur, or what other intermediate attractors may arise.

\section{Discussion and conclusions}
\label{sec:discussion}

\subsection{Limitations}
\label{sec:limitations}

Global continuation has three noteworthy limitations.
The first is the limitations of the basin map (introduced in detail in \S\ref{met:basin_maps}).
Each basin map has different limitations, and these affect directly global continuation.
For example, the \emph{recurrences-based} basin map scales poorly (in computational performance) in very high-dimensional systems of 100-1000+ dimensions.
On the other hand, the \emph{featurize-and-group} basin map does not suffer any scaling penalty as dimensions are increased. It requires the practitioner, however, to have a rough idea of what properties of the attractors would distinguish them well before running the continuation.

The second noteworthy limitation is difficulties in the presence of long transients before trajectories converge to attractors.
These can be stickiness in particular regions of the state space (such as around a ghost~\cite{Deco2012}), or the presence of transient chaos in non-chaotic attractors.
In either case, global continuation faces first a computational penalty (the larger the transients, the slower the computation), but also a conceptual penalty.
Long transients can be incorrectly classified as attractors, depending on the basin map configuration and the shortness of integration time to be skipped.
On the other hand, it may be desirable to classify transients as ``attractors'' in the context of metastable states, see \S\ref{met:basin_maps}.

The final limitation is the lack of identification of unstable sets, such as unstable limit cycles.
In the context of this article, which places emphasis on observable characteristics of a dynamical system, this is not a large limitation.
Nevertheless, unstable sets often organize the structure of the state space~\cite{Datseris2022NonlinearDynamicsJulia}, and are useful for a deeper understanding of the behaviour of the system.
Additionally, they can be useful in application contexts.
For example, unstable periodic orbits can be stabilized via feedback control~\cite{Kruse2025}.
This can be applied in a variety of contexts, from atomic force microscopy~\cite{Wallner2026} to space navigation or satellite control~\cite{Elobaid2022}.
This is why we believe global continuation is an excellent complement to local continuation and not a replacement.
Indeed, this is how we performed a detailed comparison with local continuation in \ref{sec:comparison_multiparameter}: attractors found via global continuation were used as starting points in local continuation.

\subsection{Conclusions}

This research article is the result of an appreciation and understanding of a recent shift in focus within the broader scientific community of applied dynamical systems.
This new focus is, on one hand, targeting multistable systems and critical transitions, while on the other hand placing an interest in observable and functional system behaviour that is more appropriate for realistic or high-dimensional models such as power grids, neural networks, or multispecies ecological networks.
To address this new research focus, in this paper we introduced the technique of global continuation and highlighted the plethora of its unique features, many of which cannot be achieved with alternative tooling.
In the Supplement (\ref{sisec:history}) we provide a brief accounting of how the technique of global continuation came to be.
% This is a key reason why global continuation enables fundamentally new types of analyses.
% Global continuation also effortlessly lends itself to the study of multiple, and even uncertain, parameters that are persistently present in complex models.

We close by stressing one of the most important aspects of global continuation: its simplicity and accessibility (with details in \S\ref{sec:comparison}).
Due to this, global continuation can be utilised by a wide spectrum of researchers who could benefit from dynamical systems analysis, without requiring them to be experts in numerical continuation.
From ecologists to climate scientists to power grid analysts and more, we believe global continuation will pave the way for dynamical systems analysis to be utilised and lead to new findings in a much larger scientific community.
This will be further enhanced by its professional and modern software implementation in DynamicalSystems.jl, allowing the method to be used with only a handful of transparent lines of code, as shown in Methods \S\ref{met:code}.

\section{Methods}

\subsection{Basin maps}
\label{met:basin_maps}

A basin map is a function $\mathcal{B}$ that maps initial conditions to unique integers (i.e., their basins) under the flow of the system $\Phi$. What is ultimately desired from $\mathcal{B}$ is an appropriate classification of initial conditions or perturbations from a reference state.
In the majority of cases, this is a classification into formally attracting invariant sets.
But, for example, for stochastic systems the concept of converging into an attractor is not as trivial, while the concept of the basin map applies more naturally.
Moreover, sometimes what is actually desired is converging to a particular state of operation, and this can be achieved already before rigorous convergence to an attracting invariant set.

Various different basin maps can be created, some specialized to specific types of dynamical systems such as steady states of nonlinear oscillators~\cite{HarmonicBalance}.
A key breakthrough that enables global continuation is the recent developments of fully generic basin maps that apply to all kinds of dynamical systems, with various degrees of prior knowledge required.
We will summarise three such basin maps here that are also available out of the box via their software implementation in DynamicalSystems.jl.

The first basin map is the \emph{recurrences-based} implemented in the software as \verb|BasinMapRecurrences|. Algorithmically, this is the most complex basin map, and its details are described in Ref.~\cite{Datseris2022BasinsAttraction}. Conceptually, the basin map relies on the Poincar\'e recurrence theorem~\cite{Datseris2022NonlinearDynamicsJulia}: a trajectory that has converged on an attractor will revisit every state space point it has visited before arbitrarily closely.
A smart algorithm counts accumulating recurrences by tessellating the state space into a grid of finite size, marking visited cells as a trajectory is evolved.
A finite-state machine defined on top of that grid can then understand how different initial conditions converge to different attractors, and thus find those attractors along with their basins.
The \emph{recurrences-based} basin map is incredibly powerful because it does not require any prior knowledge of the system, only an arbitrarily large region that may contain attractors.
The metaparameters of the algorithm, however, which relate to the tessellated cell sizes, how many accumulated recurrences to count for, and other aspects, are non-trivial to optimize and can affect the computational performance of the algorithm.
The main disadvantage of this basin map is that it scales as  $\tau \cdot (1/\varepsilon)^\Delta$ with $\tau$ the average recurrence time or period of an attractor, $\Delta$ its fractal dimension, and $\varepsilon$ the input tessellation size.
As such it can become very slow for high dimensional systems.

The second basin map is the \emph{featurize-and-group} implemented in the software as \verb|BasinMapFeaturizeGroup|.
It relies on the idea that trajectories that correspond to the same operating state of a dynamical system (e.g., an attractor) must share similar \emph{features}.
These are statistics extracted from the timeseries that describe various aspects of the trajectory, such as means, standard deviations, temporal variability, or other quantities.
Each initial condition is evolved for a set amount of transient and recording time, and is then mapped into a vector of features.
All the vectors of features from different initial conditions are then grouped into unique groups, using, for example, a clustering algorithm.
The unique groups are assumed to correspond to the unique attractors or unique states of operation of the system.
This basin map is very flexible because it allows for a multitude of different algorithms for how to group the features.
For example, one can prescribe predefined templates for features, and initial conditions are mapped to the template which is closest in feature space.
Another possibility is to group features using the Density-Based Spatial Clustering of Applications with Noise algorithm (DBSCAN), which is also the default option.
An alternative algorithm similar to DBSCAN but much faster and less precise is also provided.
Lastly, features can be grouped by mapping them to the cells of a predefined tessellation (histogram) of the feature space.
The featurize-and-group basin map is also suitable for mapping initial conditions to metastable states by choosing appropriate transient and recording times.
The main disadvantage of this basin map is that the practitioner must decide a-priori which features can separate the different operating states, which requires some trial and error.
In addition, the ``optimal'' features may change during a continuation, further complicating things.
Nevertheless, it is the basin map of choice for very high dimensional systems and e.g., in Ref.~\cite{Datseris2026Intermingled} it was successfully applied to spatiotemporal climate simulations that have millions of degrees of freedom.

More detailed comparison between the recurrences-based and featurize-and-group method is provided in Ref.~\cite{Datseris2023FrameworkGlobalStability}, showcasing how each has unique advantages and can be beneficial to use for different systems.
Note that in this work, we have improved the global continuation algorithm ASCM to work with both recurrences-based and featurize-and-group basin maps, which was not the case in Ref.~\cite{Datseris2023FrameworkGlobalStability}.
An alternative global continuation algorithm called FGAP is discussed in the Supplementary Information in \S\ref{sisec:history}.

The last basin map we describe here is \emph{proximity-based} implemented in the software as \verb|BasinMapProximity|.
A predetermined number of sets in the state space is provided, each with its own integer ID. Most often, these sets are already-found attractors.
An initial condition is evolved under $\Phi$ until it reaches $\varepsilon$-close to one of the predetermined sets, in which case it is mapped to the corresponding ID.
A maximum number of iterations is prescribed, so that initial conditions that do not reach any set within this number get assigned a special ID corresponding to divergence.
The main disadvantage of this basin map is that it cannot discover new attractors or operating states. It is used primarily for refining analyses.

\subsection{Matching attractors during global continuation}
\label{met:matching}

Matching is a key component of global continuation, ensuring the continuity of the tracked attractors and their properties. In this section we will use the term ``attractors'' to refer to the continued and matched sets, keeping in mind that they may not represent true attractors in the mathematical sense.
Regardless, matching is performed during the continuation, but it can also be performed retroactively after a continuation to explore different matching configurations.
Matching is flexible and different algorithms can be chosen by the practitioner, or new ones can be created. An advanced form of matching matches attractors whose points lie in the basin of attraction of an attractor existing at the previous parameter(s). The interested reader can find the details of this advanced matching in the software documentation under the name \verb|MatchByBasinEnclosure|, among other matching algorithms.
Here we will describe the default matching algorithm that matches by set distance.
The practitioner optionally provides a measure of similarity, formally a distance $\chi(A,B)\ge 0$ between sets in the state space $A, B$ (here the found attractors). By default, $\chi$ is the Euclidean distance of the centroids of $A, B$. But it can be anything that is relevant to the practitioner. It can measure similarity by subtracting the periods of two attractors, or their maximum Lyapunov exponents.
% Or, it can be a composite function that first distinguishes by the type of the attractor. E.g., $\chi$ between a fixed point and an extensive set can be $\infty$, but if the sets are of the same type, it falls back to a Hausdorff or a centroid distance.

Matching in this algorithm is done as follows. Given $\chi$, in each continuation step, the distance $\chi_{i,j}$ between all attractors of the previous step $A_i$ and attractors of the current step $B_j$ is computed.
A threshold value $r$ can be provided, so that all values of $\chi > r$ are set to $\infty$ (so that attractors that are too dissimilar are never matched).
From $\chi_{i,j}$ matching proceeds as follows. First, the pair $i-j$ which minimizes $\chi$ is matched (that is, the ID of $B_j$ is replaced by the ID of $A_i$), and then
we set $\chi_{k,j}=\infty$ and $\chi_{i,k}=\infty$ for all possible $k$.
Then, excluding $\infty$ entries, we find the next pair $i-j$ that minimizes $\chi$, and match that pair, proceeding then to replace $\chi$ entries with $\infty$ as before.
This process continues until all entries of $\chi$ are $\infty$, where the matching process is finished.
Any elements of $B_j$ that were not matched obtain a new unique integer that does not exist in the IDs of $A_j$ (the next smallest available integer).
% An alternative way to perform matching would be to implement an optimal marriage algorithm (also known as Gale-Shapley or Deferred Acceptance Algorithm) to act on $\chi_{i,j}$.

When applied throughout the continuation, the practitioner has one additional option for matching: whether to store also in memory attractors that have disappeared in prior steps, and compare (and potentially match) against them if yes.
This is particularly useful when performing continuation along arbitrary curves instead of sequential increase of a single parameter, as is the case in Fig.~\ref{fig:hilbert_continuation}, or if performing continuation along closed loops in parameter space.

\subsection{Available quantifiers}
\label{met:quantifiers}
Global continuation can track any scalar or vector-valued function of the identified attractors and/or their sampled basins. Thus, the quantities reported along a parameter curve are not limited to basin fractions: users may define arbitrary custom quantifiers and evaluate them during the continuation. Out of the box, the broader \texttt{DynamicalSystems.jl} ecosystem provides standard attractor-based diagnostics such as Lyapunov exponents and fractal dimensions, while the global continuation functionality calculates for free the basin- and resilience-based quantifiers summarised in Table~\ref{tab:available_quantifiers}; see also Ref.~\cite{Morr2026}.

\begin{table*}[t]
\centering
\small
\renewcommand{\arraystretch}{1.2}
\setlength{\tabcolsep}{6pt}
\begin{tabular}{p{0.2\textwidth}p{0.72\textwidth}}
\toprule
\textbf{Quantifier} & \textbf{Description} \\
\midrule
\multicolumn{2}{l}{\textit{Local stability}} \\
\midrule
Characteristic return time
& Slowest asymptotic convergence rate in the linear regime of a point attractor, computed from the largest real part of any eigenvalue of the Jacobian matrix as $-1/\lambda_\mathrm{max}$. \cite{May1973StabilityEcology, Pimm1984Resilience, Holling1996EngineeringRV, Scheffer2009EWSNature, Arnoldi2016ResilienceMathComparison, Smith2022VegetationResilience} \\

Reactivity
& Largest immediate response of the linearized system to a unit disturbance away from a point attractor, measured as the initial derivative of the trajectory norm. \cite{Neubert1997AlternativeResilience, Verdy2008Reactivity, Snyder2010Reactivity, Arnoldi2016ResilienceMathComparison, Liu2021Reactivity, Capdevila2020Resilience} \\

Maximal amplification and amplification time
& Largest finite-time excursion away from a point attractor among all unit disturbances in the linear regime, together with the time at which this excursion is attained. \cite{Neubert1997AlternativeResilience, Arnoldi2016ResilienceMathComparison, Liu2021Reactivity, Capdevila2020Resilience} \\

\midrule
\multicolumn{2}{l}{\textit{Nonlocal stability: geometrical measures}} \\
\midrule
Minimal critical shock
& Smallest distance from the attractor to any point on the boundary of its basin of attraction. \cite{Peterson1998Resilience, Beisner2003ResilienceLakes, Kerswell2014ResilienceOptApproach, Klinshov2015DistanceThresholdNumerical, Ashwin2016NetworkThresholds, Klinshov2020SwitchingThreshold, Halekotte2020MinimalFatalShocks} \\

Maximal non-critical shock
& Largest distance from the attractor to a point that still belongs to its own basin of attraction. Introduced in Ref.~\cite{Morr2026} \\

Basin stability
& Probability mass, or volume, of the basin of attraction under a chosen distribution of initial conditions. \cite{Holling1973ResilienceStabilityEcological, Gruemm1976DefinitionsResilience, Menck2013BasinStability, Lundstrom2018FindNonlocalResilience} \\

Intermingledness
& Ratio of average intra-group distance to average inter-group distance for labelled attractors or basins, evaluated along chosen diagnostic variables or feature dimensions. Values close to one indicate that the corresponding attractor or basin is strongly mixed with others along that diagnostic. \cite{Datseris2026Intermingled} \\

\midrule
\multicolumn{2}{l}{\textit{Nonlocal stability: transient measures}} \\
\midrule
Convergence time
& Time required for an initial condition to reach an $\varepsilon$-neighbourhood of the attractor; values over a basin can be summarized by their mean, maximum, or median. \cite{ONeill1976ResilienceEcosystem, DeAngelis1989ResilienceNutrients, Neubert1997AlternativeResilience, Cottingham1994Resilience} \\

Convergence pace
& Convergence time normalized by the distance between the initial condition and the attractor; basin-level values can again be summarized by their mean, maximum, or median. Introduced in Ref.~\cite{Morr2026} \\

Finite-time basin stability
& Basin stability restricted to the finite-time basin, i.e., to initial conditions that reach an $\varepsilon$-neighbourhood of the attractor before a prescribed time horizon. \cite{Lundstrom2018FindNonlocalResilience, Schultz2018ShockFirstExitTime} \\

\midrule
\multicolumn{2}{l}{\textit{Additional quantifiers (notable out-of-the-box examples available in DynamicalSystems.jl)*}} \\
\midrule
Lyapunov exponents
& Asymptotic exponential rates of separation of nearby trajectories. They quantify local instability of an attractor and can be tracked as scalar or vector-valued attractor diagnostics. \cite{Datseris2022NonlinearDynamicsJulia} \\

Fractal dimensions
& Dimension estimates that quantify the effective geometrical complexity of an attractor or invariant set, for example via box-counting, correlation, or extreme value theory. \cite{Datseris2023Fractal} \\

Complexity measures
& Quantifiers of the complexity in temporal variability of a timeseries.
% They are typically defined so that more complex variability yields higher complexity measure.
Examples include permutation or sample entropy~\cite{ComplexityMeasures.jl}, or recurrence quantification analysis~\cite{RecurrenceMicrostatesAnalysis.jl}. \\
\bottomrule
\end{tabular}
\caption{Available resilience-related quantifiers that can be tracked during global continuation. Local quantities are based on the linearized dynamics near fixed points, whereas nonlocal quantities use sampled basin geometry or transient behaviour.
*DynamicalSystems.jl provides implementations for hundreds of quantifiers utilized in dynamical systems applications.
Because everything is implemented in a single cohesive ecosystem, all of these quantifiers can be tracked during the continuation with minimal effort from the practitioner (couple of lines of code).
Note: these additional quantifiers are not computed for free, while the rest are.
}
\label{tab:available_quantifiers}
\end{table*}

\subsection{Code example}
\label{met:code}

The code snippet listed below is a fully runnable Julia code that recreates exactly the continuation of Fig.~\ref{fig:hilbert_continuation}b.
The snippet does not require any additional setup whatsoever (besides installing the Julia language), as even the installation of the DynamicalSystems.jl library happens during the code execution. Plotting is not included in the code snippet.

\lstset{
  literate={ε}{$\varepsilon$}1
           {Δ}{$\Delta$}1
}
\begin{lstlisting}[language=Python]
# install and load package (run on Julia v1.12)
import Pkg; Pkg.add("DynamicalSystems")
using DynamicalSystems

# define dynamical system
function predator_prey(u, p, t)
    A, B, C, D, E = p
    x, y = u
    s = x^2/(A*x^2 + B*x + 1)
    dx = x*(1 - x)*(x - E) - s*y
    dy = y*(C*s - D)
    return SVector(dx, dy)
end
p = [2.05, -2.6, 0.4, 1.0, 0.4]
u0 = [0.5, 0.02]
ds = CoupledODEs(predator_prey, u0, p)

# Define basin map (featurize-and-group)
using Statistics: mean
function featurizer(A, t)
     SVector(mean(A[:, 1]), mean(A[:, 2])/0.05)
end
gconfig = GroupViaPairwiseComparison(threshold = 0.25)
mapper = BasinMapFeaturizeGroup(
    ppds, featurizer, gconfig;
    Ttr = 1000, T = 200.0, Δt = 1.0)

# Define Hilbert 2D parameter curve (3, 5 = param. indices)
pcurve_spec = [
    3 => (0.35, 0.45, 2^3), 5 => (0.3, 0.5, 2^3)]
pcurve = hilbert_pcurve(pcurve_spec)

# run global continuation
accumulator = StabilityMeasuresAccumulator(mapper)
ascm = AttractorSeedContinueMatch(accumulator)
xg = range(0.01, 1.1; length = 11)
yg = range(0.001, 0.1; length = 11)
S = [[x, y] for x in xg for y in yg] # sampling
qcont, acont = global_continuation(ascm, pcurve, S)

# easy processing: find all parameters' values
# that permit a coexistence state
function has_coexistence(attractors)
    return any(A -> mean(A[:, 1]) * mean(A[:, 2]) > 1e-6,
        values(attractors))
end
pidxs = findall(has_coexistence, acont)

# or obtain the minimal critical shocks for all
# existing attractors during continuation:
all_mcs = qcont["minimal_critical_shock_magnitude"]
\end{lstlisting}

Following from this code snippet, it is straightforward to aggregate attractors into those that species 2 exists or not (i.e., coexistence state), and obtain the corresponding aggregated fractions (this is also what is displayed in example \ref{ex:bayesian}):

\begin{lstlisting}[language=Python]
# re-use `has_coexistence`
featurizer(A) = Int[has_coexistence(A)] # must be vec.
templates = Dict(1 => [1], 2 => [0])
gconfig =  GroupViaNearestFeature(templates)

agg_acont, _, members_cont = aggregate_continuation(
    acont, featurizer, gconfig)

agg_fcont = aggregate_fractions(
    qcont["basin_fraction"], members_cont)
\end{lstlisting}

\subsection{Open research}
\label{met:open}

The code used to create all figures in this article is open access and fully reproducible. It is available on GitHub at \url{github.com/Datseris/GlobalContinuationPaper} and archived on Zenodo, \cite{paper_codebase}.

\section*{Acknowledgements}
We thank Joseph Paez Chavez for help implementing the vibro-impact system used in example \ref{ex:engineering}.
This is ClimTip contribution \#174; the ClimTip project has received funding from the European Union's Horizon Europe research and innovation programme under grant agreement No. 101137601.

% Comment out the following to exclude SI from main document:
% (SI must be part of document for arXiv)
\onecolumn % Switch to one-column mode
\begin{center}
    \vspace{2cm} % Add vertical space
    {\Large \textbf{Supplementary Information}}
\end{center}
\twocolumn % Switch back to two-column mode
\renewcommand{\thesection}{SI\arabic{section}}
\setcounter{section}{0} % next section will be 7
\renewcommand{\thefigure}{SI\arabic{figure}}
\setcounter{figure}{0}

\section{Additional examples of global continuation applications}

%\subsection{Supports chaotic attractors}
%\label{ex:chaos}

%Deterministic chaos, while prevalent in many real world processes, is difficult to study with existing standardized tools such as local continuation. The existence of chaos must be manually inferred in some way or another in tandem to a bifurcation diagram. If the system transitions to chaotic behaviour, and a chaotic attractor is the only attractor in the state space, its identification can be done by computing in parallel to the continuation an indicator of chaos such as Lyapunov exponents for an arbitrary initial condition. Lyapunov exponents are not part of local continuation software however, so they need to be computed externally by the practitioner. Regardless, when multiple alternative attractors co-exist, and some are chaotic, this method fails and manual intervention is all that is left.

% For the first time ever to our knowledge, global continuation provides a standardized tool that can detect, and continue, chaotic attractors and identify when they appear or disappear without manual interventions. This is possible because (1) the basin map $\mathcal{B}$ works irrespectively of the attractor type an initial condition ends up to, and (2) the global continuation automatically provides, for each co-existing attractor individually, various quantities that detect chaos, such as Lyapunov exponents or fractal dimensions.

\subsection{Continuation of various properties along with attractors in a chaotic model}
\label{ex:lorenz84}

Global continuation can also continue, alongside the attractors and their basin fractions, additional properties that quantify the attractors or their basins of attraction.
These properties can be measures of resilience, or nonlocal stability, as summarized in \cite{Morr2026}. But they can also be arbitrary properties of the attractors or their basins, such as the Lyapunov exponents, extent in state space, or whatever else the practitioner is interested in.
Morr et al~\cite{Morr2026} show how this is a unique advantage of global continuation that can help the broader community to identify precursors (also called early warning signals) of tipping.

\begin{figure*}
    \centering
    \includegraphics[width=\linewidth]{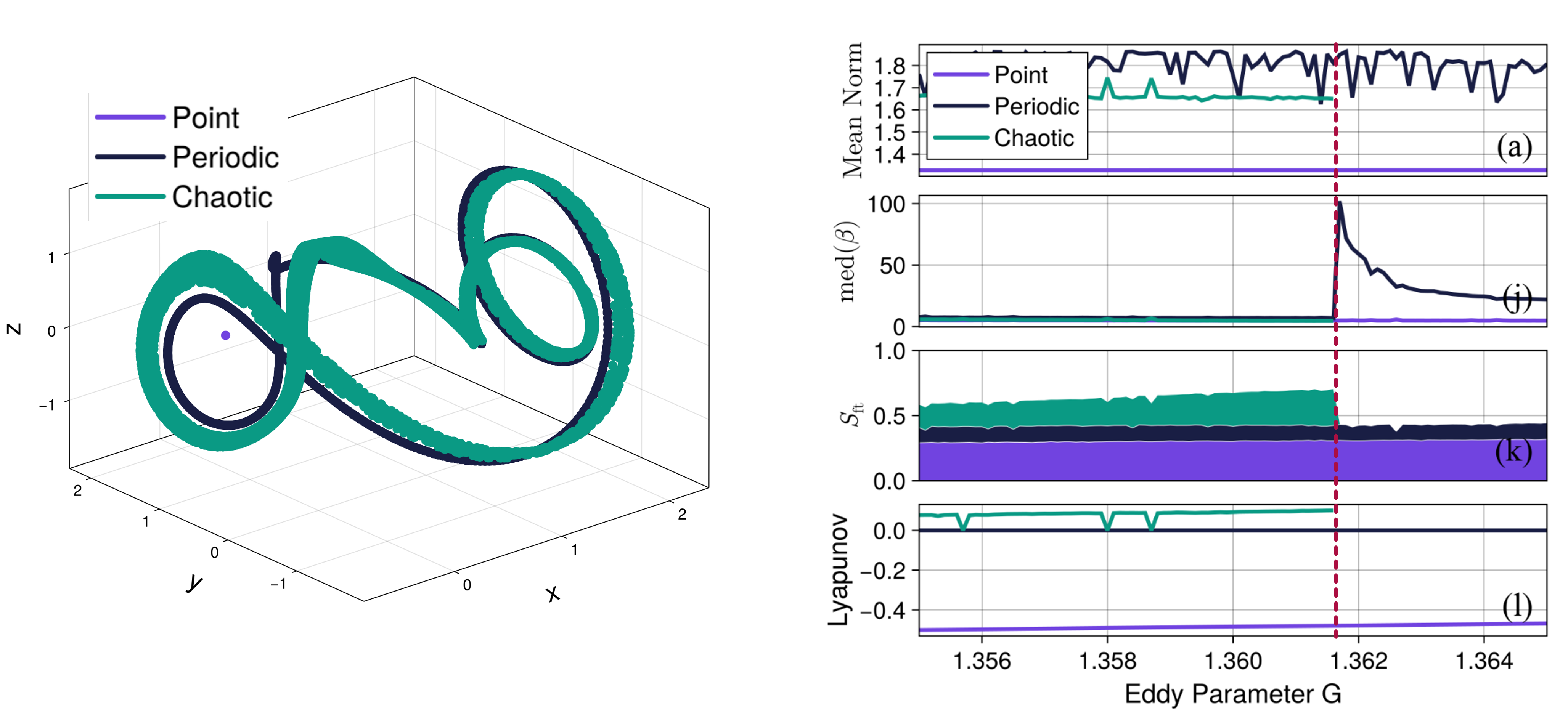}
    \caption{
    \textbf{Global continuation of a chaotic multistable model (Lorenz84)}.
    Left panel: attractors for $G=1.43$. Right panel: global continuation result, continuing (a) the attractors (and plotting their mean norm), (b) the median convergence pace, (c) the finite-time basin stability, and (d) the maximum Lyapunov exponent.}
    \label{fig:lorenz84}
\end{figure*}

To demonstrate this, in Figure~\ref{fig:lorenz84} we show the global continuation of the Lorenz-84 model~\cite{Lorenz84}. The global continuation is enhanced (automatically, practically without any effort from the practitioner) with three additional quantifiers. 
The maximum Lyapunov exponent is a well known quantity that is often used to identify chaotic behaviour~\cite{Datseris2022NonlinearDynamicsJulia}.
The convergence pace is defined as the time a perturbation takes to converge back to an attractor, divided by the magnitude of the perturbation. It is an indicator of how quickly the system stabilizes back to a particular state of operation. The median averages the convergence pace across all perturbations that return to the same attractor. Lastly, the finite-time basin stability captures the proportion of perturbations that converge to a particular attractor within a finite time window $T$, see \cite{Morr2026} for details.

The vertical dashed line in Fig.~\ref{fig:lorenz84} denotes a bifurcation where (when seen from right to left) a new chaotic attractor emerges. In the context of the Lorenz-84 model this conveys the emergence of large scale turbulence in the atmospheric flow.
We can see clearly that prior to this bifurcation, the convergence pace of an already existing limit cycle attractor increases.
This is can be clearly interpreted as an early warning signal, and the reason for this increase further analysed.
Indeed, subsequent analysis shows that orbits get stuck in the channel of a saddle-node bifurcation, spending the majority of their time there, before ultimately converging to the limit cycle (not shown). 
If the location of the saddle node in state space has clear observable characteristics, this can be the early warning signal for persistent chaos.

\subsection{Using (aggregated) basin fractions for Bayesian parameter inference}
\label{ex:bayesian}
In any realistic setting, dynamical system modelling is partially clouded with uncertainty. Even when assuming that the model equations adequately capture reality, the system parameters may be uncertain. 
Likely there is a range, or prior distribution, of parameter values that we have a rough estimate about, but the real system we hope to model will be operating over a more limited set of parameters. 

Dudowski and Kapitaniak~\cite{Dudkowski2024} showed how the basin fractions (or basin stability~\cite{Menck2013BasinStability}) of the system attractors can be used to reduce the uncertainty over the operating parameter space by utilizing Bayesian inference.
Let's assume a distribution $\mathrm{P}(p_*)$ representing initial uncertainty over a range of system parameters $p_*$. Let's also assume that we observe the system at a particular attractor $A_\circ$, out of the multiple that coexist. Bayes theorem implies the following
\begin{align}
    \mathrm{P}(p_* | A=A_\circ) = &  \frac{\mathrm{P}(A=A_\circ | p = p_*) \cdot \mathrm{P}(p_*)}{\mathrm{P}(A=A_\circ)} \nonumber \\
    = & \frac{F_\circ(p_*) \cdot \mathrm{P}(p_*)}{\int_p F_\circ(p) \mathrm{d}p}
    \label{eq:bayesian}
\end{align}
where $F_\circ(p_*)$ is the basin fraction of $A_\circ$ at parameter(s) $p_*$.
In Bayesian inference $\mathrm{P}(p_*)$ is called the prior and $\mathrm{P}(p_* | A=A_\circ) $ the posterior distribution, i.e., our updated uncertainty given the observation of $A_\circ$.

Here we showcase that this principle can be even more powerful than utilizing just the observation of a specific attractor $A_\circ$, by using basin aggregation.
The idea is that various attractors $A_j$ may share some properties of interest.
These attractors, or more specifically their basin fractions, can be aggregated based on these properties, so that groups of attractors that are similar become single entities.
A classic example where this is useful is in the study of power grids, where despite the coexistence of potentially hundreds of attractors, one is often interested in distinguishing between synchronous or asynchronous grid operation~\cite{Gelbrecht2020}.
Basin aggregation can be straightforwardly achieved by global continuation, because its shares the same principles as the ``featurize and group'' process that can be employed as a basin map $\mathcal{B}$. We expose this in more detail in the methods section \S\ref{met:aggregation}.

\begin{figure}
    \centering
    \includegraphics[width=\linewidth]{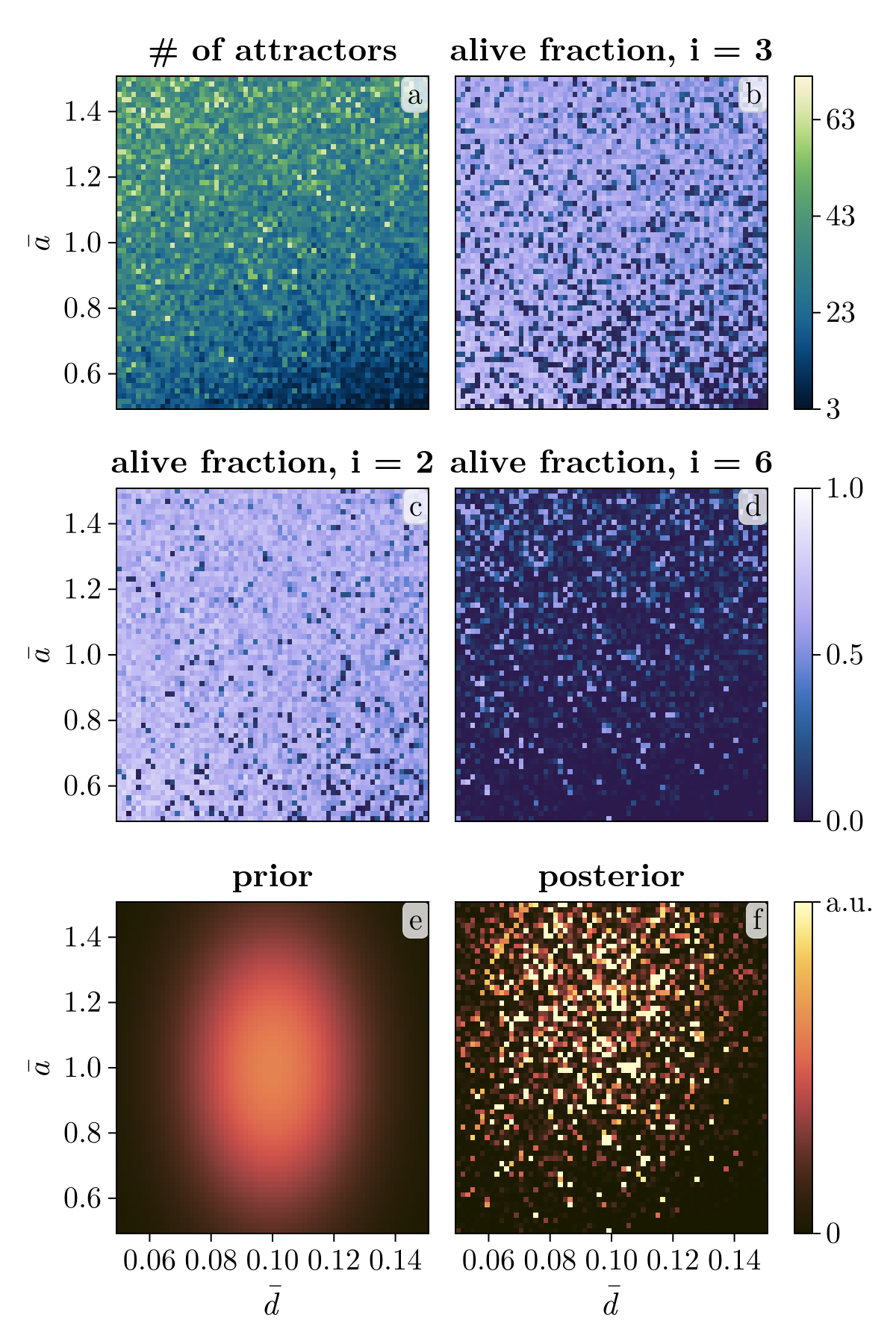}
    \caption{\textbf{Global continuation and basin fraction aggregation used for Bayesian parameter inference.}}
    \label{fig:taxonomy}
\end{figure}

To exemplify the principle, we will use a taxonomy model representing a generic ecosystem by Aguade et al.~\cite{Aguade-Gorgorio2024}. 
They used the model to show that in complex ecological communities multistability of high degree (3 or more coexisting states) is the rule, not the exception.
We used a six dimensional version of the model whose details are given in SI.
Figure~\ref{fig:taxonomy}(a) shows the results of a global continuation of the model over its two-dimensional parameter space.
It becomes clear that this model is characterized by extreme multistability. 
Let's now assume that we monitor an ecosystem where Eq.~\eqref{eq:taxonomy} is applicable for. Given the extreme multistability, fitting the model parameters is not straightforward, as we may be fitting the wrong attractor.
Let's assume however that we observe that species 2, 3 and 6 are alive.
For each species $i$ we perform an aggregation of basin fractions: at each parameter, all attractors for which species $i$ is not extinct are aggregated. This removes the difficulties of focusing on a particular attractor and allows us to focus on the quantities we can actually observe.
The aggregated fractions are presented in Figure~\ref{fig:taxonomy}(b-d). 

From the single data point of ``species 2, 3, 6 exist'' we have now generated three independent observations (parameter probability density functions) that can be passed through Eq.~\eqref{eq:bayesian} to update our prior. 
The final posterior distribution is shown in Fig.~\ref{fig:taxonomy}(f).
We expect this to be a valuable tool for parameter inference, allowing one to prepare a dynamical-systems-informed prior that can the utilized in more extensive parameter fitting endeavours.

This process of attractor aggregation goes well beyond simply aggregating their basin fractions that we showcased here. The entire basins, and their derived properties such as convergence times, can be aggregated, something that we highlight in example \ref{ex:aggregation}.

\subsection{Aggregating arbitrary continuation properties according to system operating state}
\label{ex:aggregation}

Normally attractors are separated by their location in state space, but this does not necessarily correspond to the system observed characteristics or operating states. A typical example are synchronized states in power grid operation, which can occur in a plethora of different attractors existing in different state space locations. Aggregation allows one to focus on the qualities of the system that matter for its real world application.

A further, practically important extension is that the properties being continued need not refer to individual attractors.
In many applications the practitioner cares about the stability of a mode of operation that spans several distinct attractors rather than any one of them in isolation.
Schoenmakers and Feudel~\cite{Schoenmakers2021FunctioningResilience} formalise this through the notion of system functioning: a system is functioning, e.g., as long as its state falls within some prescribed subset of state space, regardless of which specific attractor it resides on. Global continuation can accommodate this directly. Attractors that share the same mode of functioning are aggregated into a single entity (as described in \S\ref{met:aggregation}) and all stability quantifiers are then accumulated over the combined basin, yielding measures that describe the resilience of the functioning mode as a whole.

\begin{figure}
    \centering
    \includegraphics[width=\linewidth]{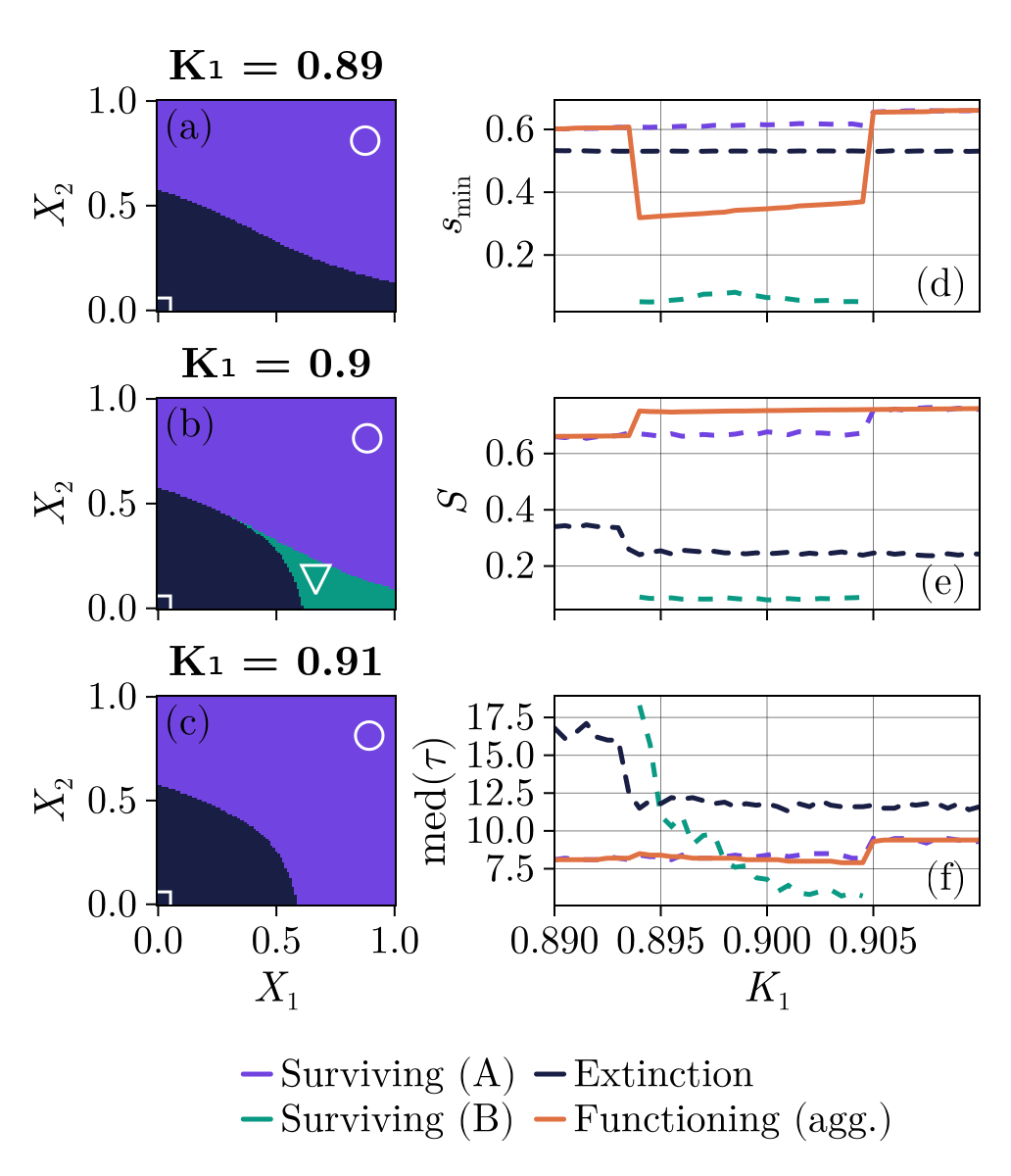}
    \caption{\textbf{Global continuation of an aggregated functioning mode with aggregated stability.}
    (a--c) Basins of attraction of the two-habitat population model at three values of the carrying capacity $K_1$, bracketing the bistable window.
    Violet: high-biomass surviving state; teal: low-biomass surviving state; dark blue: extinction.
    (d--f) Stability quantifiers along the continuation:
    minimum critical shock magnitude, basin stability, and median convergence time.
    Dashed lines show individual attractors; the solid orange line shows the aggregated functioning mode.}
    \label{fig:aggregated_stability}
\end{figure}

As an example, we apply this to the two-habitat population model with Allee effects from~\cite{Schoenmakers2021FunctioningResilience}.
As the carrying capacity $K_1$ decreases, representing habitat degradation, the system passes through a bistable window in which two distinct surviving equilibria coexist alongside an extinction state (Fig.~\ref{fig:aggregated_stability}a--c). Ecologically, the relevant question is not which surviving state the population occupies, but whether it survives at all. Aggregating both surviving attractors into a single functioning mode and continuing the stability measures over $K_1$ (Fig.~\ref{fig:aggregated_stability}d--f) gives a coherent picture of how resilient survival is as a whole, one that remains well-defined and smoothly varying even across the bistable window where individual attractor counts change. This example illustrates a general principle: by choosing what to aggregate, the practitioner decides what stability is being measured, and global continuation then tracks that stability automatically.

\subsection{Global continuation works under any kind of local or global bifurcations without intervention}
\label{ex:lossy_network}

A perfect example to highlight this feature is an oscillator network representing a power grid. 
In Ref.~\cite{Hellmann2020} the authors introduce a power grid model with realistic sparse coupling and physical losses between nodes, given by (repeating Eq. (1) of \cite{Hellmann2020})
\begin{equation}
    \ddot{\phi} = p_i - d_i \dot{\phi}_i  + \sum_{j=1}^{N} K_{ij}\left( \sin(\alpha) + \sin(\phi_i - \phi_j - \alpha) \right)
    \label{eq:lossy_network}
\end{equation}
where $p$ represents power production or consumption, $d$ is damping, and $\alpha$ represents physical loss. 
In Ref.~\cite{Hellmann2020} the authors report that this model, largely independently of network configuration, displays strong multistability and a plethora of unique states, termed solitary and exotic solitary. 
Detailed analysis shows these states to be a result of a multitude of global bifurcations.
Due to the high dimensionality of the model (400-500), the presence of a large number of local and global bifurcations makes it incredibly difficult to analyse using traditional tools of local continuation, which would require a huge amount of interventions and re-starts.

This is not at all the case for global continuation. Here we consider a ring network of 200 nodes (as in Fig.~3 of Ref.~\cite{Hellmann2020}) for which we will perform global continuation.
To follow the original study, we will not sample initial conditions randomly in the state space.
Instead, our sampling $S$ given as an input to global continuation will be a set of perturbations done to the synchronised state, as in \cite{Hellmann2020}.
This already shows how easy it is to switch global continuation from studying basin stability to studying perturbations of a specific state.
Then, for finding attractors, we will use the featurize-and-group basin map, and group features via prescribed templates.
The six unique behaviours reported in Ref.~\cite{Hellmann2020} become templates, and each trajectory of each initial condition is analysed by the featurizer and mapped into the template that it corresponds to.
We also disabled the default matching step of global continuation for this example, as the templates are prescribed the same irrespectively of parameter.

\begin{figure}
    \centering
    \includegraphics[width=1\linewidth]{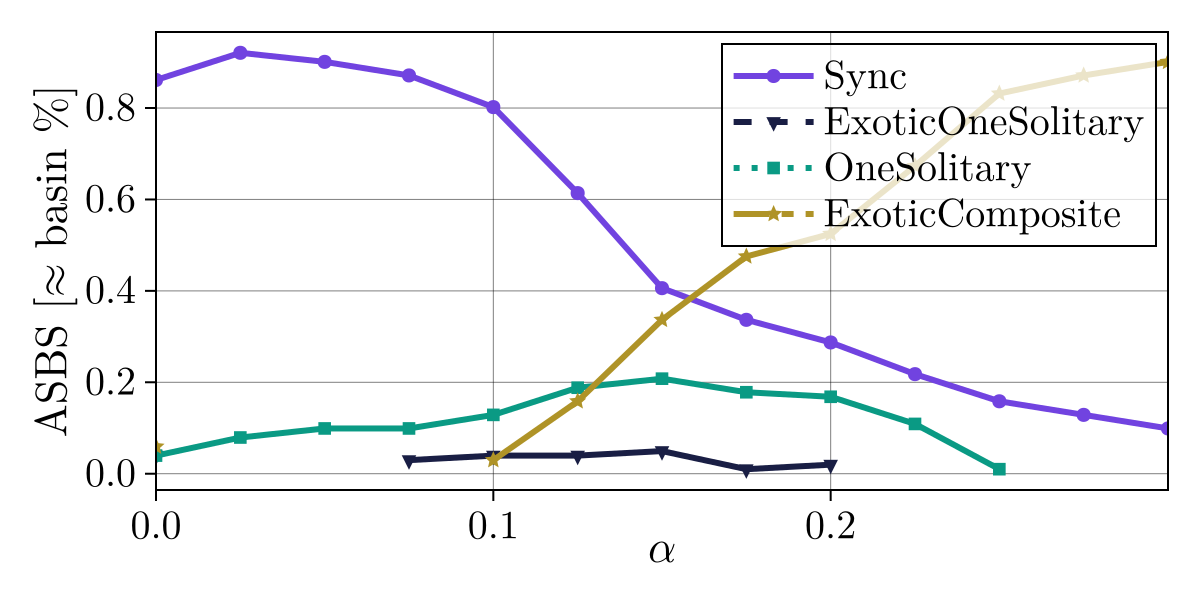}
    \caption{\textbf{Global continuation of a power grid network with loss undergoing several complex global bifurcations.}
    The result is nearly identical to Fig.~3b of Ref.~\cite{Hellmann2020}. 
    Small differences are due to our code classifying ``other'' states as exotic composites, running the analysis for substantially less number of perturbations to the sync state; and not identifying the sync state anew for each value of $\alpha$ for simplicity.}
    \label{fig:lossy_network}
\end{figure}

In Fig.~\ref{fig:lossy_network} we present the results, which came from a single run of a global continuation without interventions or reruns. 
Worth highlight is how simple it was to perform the analysis and derive at practically identical results as the original paper:
when excluding the definition of the templates and generation of perturbations (both are context specific), the actual global continuation only needed 10 lines of code (see provided codebase reference in main text).

\subsection{Global continuation of discontinuous dynamics}
\label{ex:engineering}

One of the unique features of global continuation is that it relies only on the existence of a basin map, making it straightforward to apply in discontinuous systems. 
To demonstrate this, we apply it to the vibro-impact mechanical system of Ref.~\cite{Liu2020}.
This nonautonomous dynamical system has three dynamic variables and multiple discontinuities in the vector field and in the driving force. As such, standard local continuation techniques do not apply. The authors had to implement advanced numerical techniques in the COCO continuation software to study its parameter dependence. 

\begin{figure}
    \centering
    \includegraphics[width=1\linewidth]{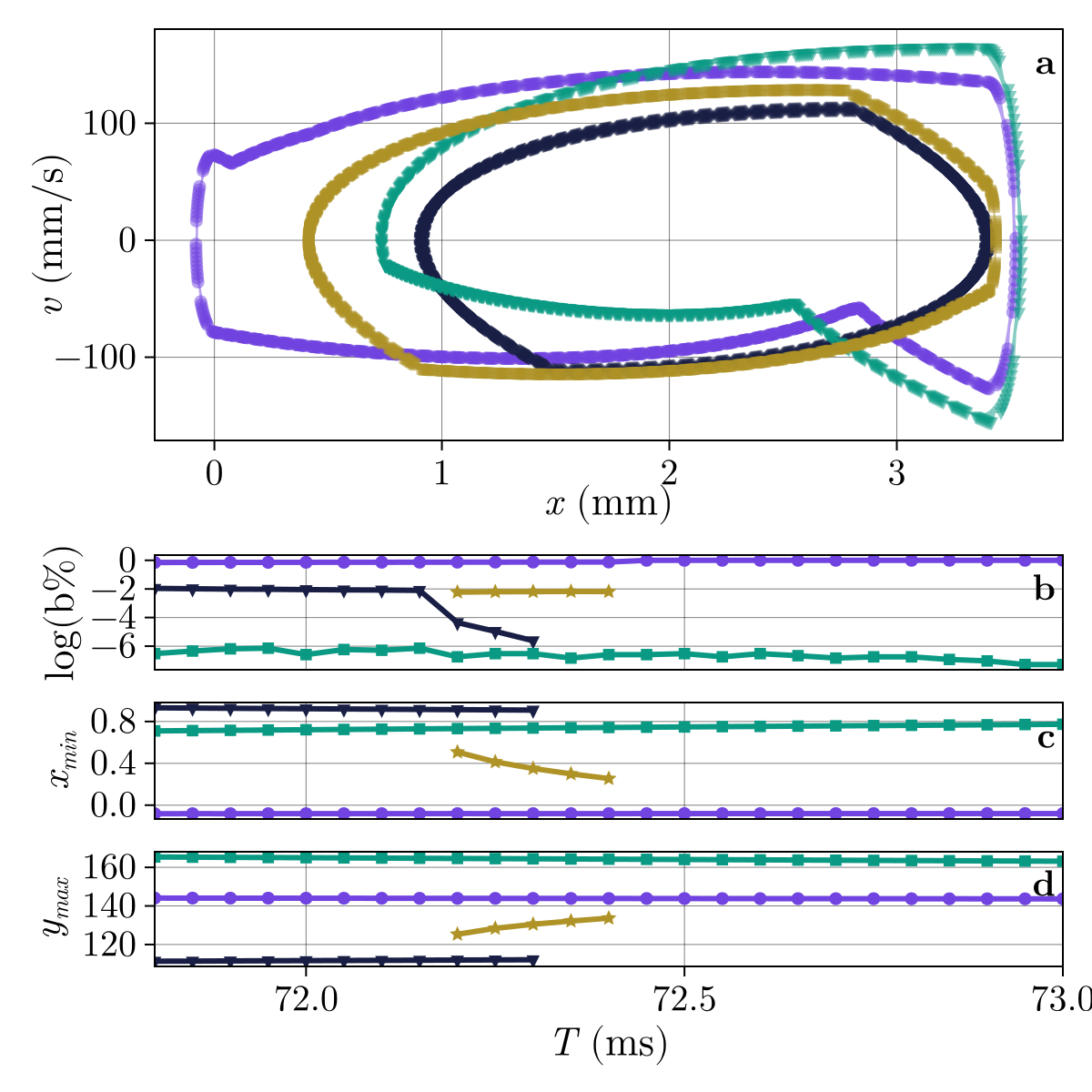}
    \caption{\textbf{Global continuation of a vibro-impact mechanical system with multiple discontinuities in vector field and driving force.} The parameter configuration of the model is the same as in Fig.5(b) of Ref.~\cite{Liu2020}. a: attractors for $T\approx72.25$. b: logarithm of basin fractions. c, d: features used in the featurize-and-group basin map.
    The purple, black and yellow attractors are also found in Ref.~\cite{Liu2020}, but not the teal/green one.}
    \label{fig:vibroimpact}
\end{figure}

Here, we do not deploy any advanced analysis.
We create a featurize-and-group basin map, and use as features simply the minimum and maximum of two of the three variables of the system.
We then run the global continuation without modifications from its default settings.
The continuation of the system is presented in Fig.~\ref{fig:vibroimpact}. Despite the lack of effort in adjusting or augmenting global continuation, we immediately replicate the results of the original article.
Not only this, but we are able to find a new attractor that was missed in the original article because (a) it has small basins of attraction, and (b) its local continuation ``branch'' never connects to the found branches in Ref.~\cite{Liu2020}. 
This highlights how useful global continuation is: not only it is effortlessly applicable to discontinuous systems, but it also does not require the practitioner to know in advance what, or where, to look for.

\section{Aggregation of global continuation}
\label{met:aggregation}

Global continuation tracks the different attractors of a dynamical system individually.
In some applied contexts, however, the practitioner may not care about every fine-grained dynamical distinction between attractors, but about broader \emph{modes of operation}: groups of attractors that share a common functional role in the context of the dynamical system. 
This is highlighted in examples \ref{ex:aggregation}, \ref{ex:bayesian}, \ref{ex:lossy_network} (note that last example presents aggregated modes but does not actively aggregate using the functionality described here).

Aggregation of global continuation is straightforward to achieve, especially with the code implementation that accompanies it in DynamicalSystems.jl.
It reuses fully the featurize-and-group basin map defined in \S\ref{met:basin_maps}.
Essentially, the practitioner prescribes the features of attractors of interest, and how to group them.
Using the same process as featurize-and-group, found attractors are aggregated (at each step of the continuation individually) and this applies also for the attractor properties: e.g., the basin fractions of aggregated attractors are summed.
Across the continuation, the aggregated attractors maintain consistent labels by employing a similar matching procedure as in \ref{met:matching}.
The only difference is that now distances are estimated between groups of features instead of attractors. Both however are sets of vectors, so the underlying mathematics and code implementation are reused.

% The population-model example (\S\ref{ex:aggregation}) illustrates the inline approach: a scalar featurizer returns $1$ if the attractor's mean total biomass falls below an extinction threshold and $0$ otherwise. \verb|GroupViaPairwiseComparison| with threshold $\delta = 0.5$ then cleanly separates extinction from surviving attractors, since their feature values differ by exactly $1$. All surviving attractors are consequently merged into a single functioning mode, and the stability measures reported for that mode (basin stability, minimal critical shock magnitude, and median convergence time) quantify the collective resilience of the species as a whole over the full continuation range.

\section{Extensive comparison between local and global continuation}
\label{sec:comparison}

We present in Table~\ref{tab:comparison} a formal and extensive comparison between local and global continuation.
One practical aspect worth discussing further is what we believe to be one of the biggest strengths of global continuation: its accessibility.
Global continuation is simple to understand and use, and does not require extensive knowledge of numerical techniques or advanced understanding of dynamical systems bifurcations. 
Essentially, the fundamental concepts of the flow $\Phi$ and the basin map $\mathcal{B}$ are all one needs to fully utilize global continuation.
Its focus on physical observables makes it even more natural for applied scientists. 
Its output is also fundamentally simple: at each parameter, the output is a (matched) set of attractors and a set of real numbers corresponding to attractor or basin properties.
Lastly, global continuation does not require the practitioner to know in advance ``what to look for'': it provides everything (all attractors and all their basin properties) all at once without manual intervention.
It also does not require different configurations or inputs when wanting different outputs, as is the case for local continuation when, e.g., searching for fixed points or limit cycles.

\newcounter{tablecounter} % Create a dedicated counter for the table, so that it is easy to comment out rows
\renewcommand{\arraystretch}{1.2}
\begin{table*}[]
    \centering
    \begin{tabular}{p{.5cm}p{9cm}p{8cm}}\toprule
    %\begin{tabularx}{\columnwidth}{ll}
    \textbf{\#} & \textbf{Local continuation (traditional bifurcation analysis)} & \textbf{Global continuation (using ASCM)} \\ \hline
    \rowcolor{blue!20} \refstepcounter{tablecounter}\arabic{tablecounter}$^\diamond$ & Finds unstable sets in the state space \newline (fixed points / limit cycles / heteroclinic structures). & Only finds attracting sets. \\
    \rowcolor{blue!20} \refstepcounter{tablecounter}\arabic{tablecounter} & Is (typically) resilient versus transients. & Long transients can be (incorrectly or deliberately) classified as attractors. \\
    \rowcolor{blue!20} \refstepcounter{tablecounter}\arabic{tablecounter}$^\times$ & Does not put limits on state space extent. & Needs as an input a state space region to search for attractors. \\
    \rowcolor{blue!20} \refstepcounter{tablecounter}\arabic{tablecounter}$^\triangle$ & Detects and classifies local (and some global) bifurcation points. & Does not explicitly detect bifurcations, but both local and global bifurcations can be inferred. \\
    \rowcolor{blue!20} \refstepcounter{tablecounter}\arabic{tablecounter} & Can be extended to handle PDEs with specialized efficiency and accuracy (e.g.,~\cite{Uecker2021}). & No special handling of PDEs. Typically slower for PDEs. \\
    \hline
    \rowcolor{green!20}\refstepcounter{tablecounter}\arabic{tablecounter}$^\odot$ & Finds and continues fixed points and periodic orbits. & Finds and continues any kind of attractors, including quasiperiodic or chaotic. \\
    %\rowcolor{green!20} \refstepcounter{tablecounter}\arabic{tablecounter} & Requires a computable Jacobian of the dynamic rule. & Does not utilize a Jacobian and can be used straightforwardly for transformed systems like Poincar\'e maps or projections. \\ 
    \rowcolor{green!20} \refstepcounter{tablecounter}\arabic{tablecounter}$^\mathparagraph$ & User must manually search for and identify multistability. & Coexisting attractors are automatically detected and returned as different objects. \\ 
    \rowcolor{green!20} \refstepcounter{tablecounter}\arabic{tablecounter}$^\wedge$ & Allows for alternative algorithms for finding limit cycles. & Allows for fundamentally different algorithms for the basin map $\mathcal{B}$. \\ 
    \rowcolor{green!20} \refstepcounter{tablecounter}\arabic{tablecounter}$^\diamond$ & Does not preserve the flow $\Phi$ or basin structure. & Preserves $\Phi$, allowing estimation of basin volumes and critical shocks. \\
    \rowcolor{green!20} \refstepcounter{tablecounter}\arabic{tablecounter} & Computes only a single and local quantifier of stability (Jacobian eigenvalues). & Computes a plethora of varied stability quantifiers, including local, nonlocal, and global. \\
    % \rowcolor{green!20} \refstepcounter{tablecounter}\arabic{tablecounter} & Uses the local, linearized dynamics. & Uses the full nonlinear dynamics. \\
    \rowcolor{green!20} \refstepcounter{tablecounter}\arabic{tablecounter}$^\ddagger$ & Limited use in indicating loss of stability. & More likely to indicate loss of stability due to the varied quantifiers of nonlocal stability reported. \\
    % \rowcolor{green!20} \refstepcounter{tablecounter}\arabic{tablecounter}$^*$ & Parameter change may not affect linear stability of all fixed points. & Parameter change is more likely to affect global stability of all attractors. \\
    \rowcolor{green!20} \refstepcounter{tablecounter}\arabic{tablecounter} & No sensible aggregation of attractors possible due to the focus on branches. & Attractors can be aggregated into sensible categories based on requested properties, during or after continuation. \\
    \rowcolor{green!20} \refstepcounter{tablecounter}\arabic{tablecounter}$^ 	\otimes$ & No flexibility on matching attractors. & Explicit and user-configurable matching of attractors. \\
    \rowcolor{green!20} \refstepcounter{tablecounter}\arabic{tablecounter} & Requires expertise and constant interventions.  & Conceptually straightforward without needing intervention \\
    \rowcolor{green!20} \refstepcounter{tablecounter}\arabic{tablecounter} & Different algorithms must be employed for multiparameter continuation. & Identical operation regardless of parameters explored. \\ 

    \hline\end{tabular}
    %\end{tabularx}
    %%%%%%%%%%%%%%%%%%%%%%%% CAPTION %%%%%%%%%%%%%%%%%
    \caption{A comparison between local and global continuation as tools for analyzing the stability of a dynamical system versus parameters. Entries are colored blue when they contain advantages of local over global, and green for the opposite. 
    %\addtabletext{
    $^\diamond$Local continuation transforms the system into a modified system where unstable states are also attracting, hence being able to find unstable sets. By definition, this makes local continuation not preserve the flow $\Phi$ and thus not preserve the structure of the basins of attraction.
    $^\times$The attractor-containing region provided for global continuation can be arbitrarily large as a first pass and refined later, so this is not a strong limitation in practice.
    $^\odot$ Some very specialized local continuation algorithms may find quasiperiodic tori, but this is not typical usage nor is available in typical software providing local continuation functionality.
    $^\triangle$Simple classification into local or global bifurcation is possible. E.g., if an attractor disappears and its basin fraction also smoothly goes to 0, this is a local bifurcation, otherwise global. But this is does not match the level of classification detail that local continuation provides.
    $^\mathparagraph$In global continuation the probability to find a new attractor is equal to $1 - (1 - f)^n$ with $f$ the basin fraction of the attractor and $n$ the amount of sampled initial conditions. In local continuation there is no known relationship between the initial condition and the point it will converge to.
    $^\wedge$This is listed as an advantage of global continuation because the different basin maps have fundamentally different pros and cons, providing more flexibility and power to the practitioner. See \S\ref{met:basin_maps} for a comparison of the basin maps.
    $^\ddagger$Changing a parameter often does not meaningfully increase the unstable eigenvalues of the Jacobian matrix, which would indicate loss of stability. On the other hand, basin fractions typically decrease smoothly towards zero as an attractor loses stability~\cite{Menck2013BasinStability}, although this is not guaranteed to be the case~\cite{Schultz2017}, in which scenario, neither method indicates loss of stability.
    $^\otimes$This is important for establishing continuity based on the physical or observable properties the practitioner cares about, see \S\ref{sec:global_cont_highlevel}.
    }
    \label{tab:comparison}
\end{table*}

\section{Comparison between local and global multiparameter continuation}
\label{sec:comparison_multiparameter}
%Local (including multiparameter) continuation is based on the implicit function theorem which states that zeroes of a continuously differentiable function 
%\begin{align} \label{eq:implicit_f}
%    f:\mathbb{R}^n \times \mathbb{R}^{k} \to \mathbb{R}^n, \: (x, p) \to f(x, p)
%\end{align}
%whose Jacobian $D_x f(x_0, p_0)$ at a point $(x_0, p_0)$ is invertible can be continued as function of the parameter $p$ near $(x_0, p_0)$ \cite{Kuznetsov2004AppliedBifurcationTheory}. Thus at a high level, local continuation is a combination of predictor-corrector Newton algorithms. Bifurcation points are identified as points where the rank of this function changes under slight changes of the parameters.

\begin{figure}
    \centering
    \includegraphics[width=0.5\textwidth]{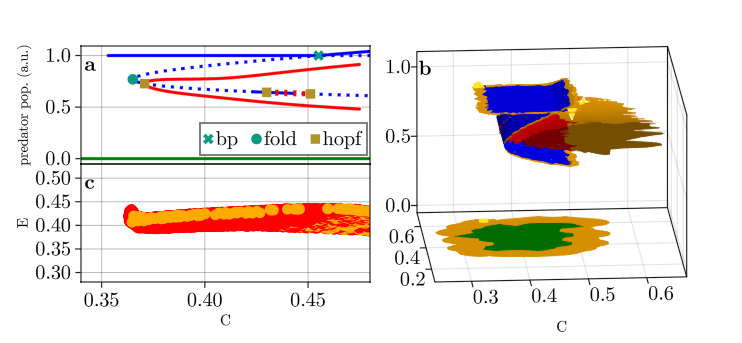}
    \caption{Local multiparameter continuation of the predator prey model using \cite{MultiParamContin} (cf.) Figure \ref{fig:hilbert_continuation}. \textbf{a} shows the bifurcation diagram with equilibria in blue and green, and periodic orbits in red (maximum and minimum). Thick and thin dashed lines indicate stable and unstable equillibria/limit cycles, respectively. \textbf{b} shows the codimension $2$ manifold of equilibria and periodic orbits (colour-matched to \textbf{a}) obtained via multiparameter continuation, with boundary charts coloured yellow. \textbf{c} is a projection onto the $(C,E)$ plane where the stable limit cycle exists. The red points correspond to interior points on the manifold of limit in \textbf{b}, and the yellow points correspond to boundary points.}
    \label{fig:multiparamcont}
\end{figure}

Figure \ref{fig:multiparamcont} shows a bifurcation diagram together with a local multiparameter continuation of all attractors that could emerge from each bifurcation for the predator-prey model used in Fig.~\ref{fig:hilbert_continuation} of the main text. Both the bifurcation diagram and the continuation were obtained using the method of Ref.~\cite{MultiParamContin} implemented in the BifurcationKit.jl software (\cite{BifurcationKit.jl} version 0.8). 
To achieve this local continuation, one has to choose appropriate initial conditions to identify each branch of equilibria, and then perform a separate multiparameter continuation for each bifurcation point over the range of parameter values of interest. 
For example, to continue the branch of periodic orbit(s) that would emerge from one of the Hopf points, we first slightly vary the parameter away from the bifurcation to find a periodic orbit. 
The function $f$ is then defined by solving the ODE up to the period, and the last point along this orbit is subtracted from the initial point on the orbit. The zeros of $f$ are then continued in the joint parameter and period space. Thus, some foreknowledge of the branches/attractors must be used to obtain initial conditions which would converge under Newton's algorithm (except in the case of the periodic orbit, where the initially found periodic orbit is used as the initial condition). This can be achieved via preliminary analysis or as in our case, by using the outputs of global continuation. This highlights one of the drawbacks of local multiparameter continuations in that the practitioner needs to know (at least at some combination of the parameter values) all attractors in the state space already (continuing points that are not already on found attractors is problematic as local continuation may not map them to the attractor they actually converge under the flow $\Phi$).

From an algorithmic point of view, it is difficult to directly compare local and global multiparameter continuations, since they are doing different things and provide different outputs.
Here we will focus the discussion on three aspects: (a) simplicity of achieving (approximately) similar outcomes, (b) the type of output and what analyses it allows for, 
(c) the computational scaling of each method versus the number of parameters and density of coverage of the parameter space. 
(a) was discussed in the preceding paragraph already, and our conclusion is that global continuation is a substantially simpler approach.
For (b), generally speaking, global continuation provides  more information about the system, as it explores the whole state space. This advantage careers over from \S\ref{sec:comparison}. The type of output remains the same as in single parameter continuation.
This is not the case in local multiparameter continuation, as its output is clearly more complex than the output of local single parameter continuation.
Generally speaking, it is a discretized hypersurface that covers the joint state-parameter space. Answering simple questions such as ``for which parameters does a coexistence stable state exists'' is also possible in local multiparameter continuation, but requires substantial and nontrivial processing of this output.

To compare (c) performance, we must note that global continuation scales exponentially with the number of parameters. 
If $k$ parameters are continued over, each covered with $n$ number of points, global continuation scales as $n^k$.
In contrast, local continuation scales linearly with the number of parameters making it suitable for very high dimensional problems. While global Hilbert continuation covers the parameter space with points on the Hilbert curve, local multiparameter continuation covers the manifold to be continued (i.e periodic orbits or fixed points) in the extended state space of the state and parameters with charts formed from polytopes (higher dimensional polygons). We find that the closest comparison to number of points on the Hilbert curve in global continuations is the total number of points on the edges of all polytopes needed to cover the manifold. In the predator-prey example, global continuation required about two to three times the number of points on the Hilbert curve to obtain approximately similar number of points in Fig.~\ref{fig:multiparamcont} \textbf{c} and Fig.~\ref{fig:hilbert_continuation} \textbf{c}.
For two parameters their performance was comparable. It is clear however that this relationship won't hold when continuing over a higher dimensional parameter space, which is why local continuation remains a useful toolbox for the study of multiple parameters.

% text by George:
% It is worth pointing out how difficult achieving any of this would be with local continuation.  First, substantial effort would be needed to simply identify the existence of a species-coexistence attractor that is also stable, as doing so in multiple parameters is not standard local continuation but requires usage of advanced techniques based on multi-parameter continuation ~\cite{MultiParamContin, dankowicz2013recipes}. Then nontrivial postprocessing would need to be done to identify the parameter regimes where this particular attractor exists, due to the fundamentally different output that local continuation provides (joint parameter-space versus system behaviour at  given parameters). Lastly, identifying safe operating spaces in terms of system response to finite perturbations, as in Fig.~\ref{fig:hilbert_continuation}d, is simply impossible using local continuation techniques and would require additional simulations by the practitioner.

\section{Historic tracing of global continuation}
\label{sisec:history}

The basis of global continuation is the basin map, and more precisely a computational algorithm for a generic, dynamical-system-agnostic basin map.
Surprisingly, despite the many decades of research in multistable systems~\cite{Pisarchik2022}, such a concept did not exist until recently. 
% Indeed, textbooks in multistability provide no information on how to actually map initial conditions to different basins, in an algorithmic and numerical manner.
A key breakthrough came in 2020 by Gelbrecht et al.~\cite{Gelbrecht2020}, in an article that introduced both a computational algorithm for a generic basin map, and the first method of global continuation (and we note that the ideas in that article have been cultivated by the same research group over several other articles such as~\cite{Hellmann2020}).
The authors created a specific version of the featurize-and-group basin map (Methods, \S\ref{met:basin_maps}) that groups features into attractors using DBSCAN. In the article the concept ``feature'' is replaced by ``pseudometric''. 
Then, they extended the featurize-and-group basin map into a continuation by grouping features from trajectories sampled at \emph{all} parameter values $p_i$.
Therefore, the similarity of the features at different parameters also establishes the continuity of the features across the parameter (this is an alternative to the matching process of the main ASCM algorithm we introduced in this research article).
The algorithm of Ref.~\cite{Gelbrecht2020} was termed `Monte Carlo Basin Bifurcation Analysis' (MCBB). It could not find and track the actual attractors of a system, due to the merging of different attractors across different parameters. But it could track the basin fraction (basin stability) of uniquely identifiable operating states of the dynamical system.

A year later a very similar algorithm was introduced in Ref.~\cite{Stender2021} termed bSTAB, without reference to MCBB. It was also an implementation of the featurize-and-group basin map.
In addition to grouping using DBSCAN, Ref.~\cite{Stender2021} also introduced the concept of grouping by providing a set of templates (as discussed in Methods~\ref{met:basin_maps}).
Similarly with MCBB, this method could not find actual attractors, but only the basin fraction (basin stability) corresponding to uniquely identifiable operating states. Unlike MCBB however, the bSTAB method could not perform a continuation unless the provided templates remained valid throughout the parameter range, something unlikely to remain true if the system undergoes any bifurcations within this range.

Both aforementioned methods require some knowledge of the system characteristics for extracting features useful in distinguishing different operating states. In Ref.~\cite{Datseris2022BasinsAttraction} we introduced the recurrences-based basin map (Methods \ref{met:basin_maps}), which works generically for any dynamical system and any type of attractor(s), requiring only a region in state space to search for attractors as a key input. This algorithm could identify exactly individual attractors, and thus also basins of attraction, by consequence allowing the estimation of various quantities such as basin stability, or minimal critical perturbations.

Ref.~\cite{Datseris2022BasinsAttraction} however was not a continuation method. 
In Ref.~\cite{Datseris2023FrameworkGlobalStability} we extended the algorithm into a continuation, by introducing the concept of matching of attractors and random sampling. That algorithm was a precursor of ASCM. It could track the actual system attractors, and their basin fractions (basin stability) across a parameter range.
In the same article we introduced an alternative global continuation algorithm we termed FGAP (Featurize and Group Across Parameters). It is a generalization of MCBB but flexible so that it can work with any method of grouping features, including the templates method of Ref.~\cite{Stender2021}.
Later, in Ref.~\cite{Morr2026} we extended the work of Ref.~\cite{Datseris2023FrameworkGlobalStability}, so that it can track various different quantifiers of stability, such as convergence rate.

In this article, we have combined and substantially improved all prior work with the ASCM algorithm. It allows using any basin map, including future ones, during a global continuation and combines tracking and matching attractors, basin fractions, and other stability measures when possible.
We further introduced continuation over Hilbert curves to efficiently cover multi-dimensional parameter spaces while still having a sensible matching procedure. 
We implemented aggregation to also work over all these possibilities, allowing one to aggregate attractors, their basin fractions, and other stability properties, according to desired characteristics, irrespectively of the basin map used.
Finally, here we provide a coherent overview of the method and a plethora of appealing examples showcasing its usefulness in varied scenarios.

\section{Details of used models}
\label{sec:model_details}

\paragraph{Neural mass model.} The neural mass model used in Fig.~\ref{fig:comparison} comes from Ref.~\cite{Cortes2013} and is given by
\begin{align*}
   \tau \dot{E} &= -E + g(JuxE + E_0) \\
   \dot{x} &= (1-x)/\tau_D - uEx \\ 
    \dot{u} &= U_0 E(1-u) - (u - U_0)/\tau_F \\
    g(y) & := \alpha \log(1+\exp(y/\alpha))
\end{align*}
with $\alpha = 1.5, \tau = 0.013, J = 3.07, \tau_D = 0.200, U_0 = 0.3, \tau_F = 1.5$ and $E_0$ the model parameter varied as in Fig.~\ref{fig:comparison}.

\paragraph{Echo state networks.} For the example of \S\ref{ex:rnn}, the state update equation of an echo state network with $N$ hidden units, input size $m$ and output size $k$ takes the form
\begin{align*}
    x_{n+1} &= \tanh(Wx_n + W^\text{in} u_{n+1} + W^\text{fb} y_n), \\
    y_{n+1} &= W^\text{out} x_{n+1}
\end{align*}
where $W \in \mathbb{R}^{N \times N}, \: W^\text{in} \in \mathbb{R}^{N \times m}, W^\text{fb} \in \mathbb{R}^{N \times k}$ and $W^\text{out} \in \mathbb{R}^{k \times m}$ are matrices. ESNs are distinguished by the random generation of the matrices $W, \: W^\text{in}$ and $ W^\text{fb}$ which are then kept fixed during training while $W^\text{out}$ is learned via regularised regression. In Figure \ref{fig:highlight_rnn}, we used $N = 500$ and $m = k = 2$. The entries of $W, \: W^\text{in}$ and $ W^\text{fb}$ were sampled from a uniform distribution between $-1$ and $1$ and $W$ is rescaled to have the specified spectral radius in the figure. $W^\text{out}$ was found via regularised regression with teacher forced feedback. Once $W^\text{out}$ is found, the autonomous dynamics reduces to $x_{n+1} = \tanh((W + W^\text{fb}W^\text{out})x_n)$.

\paragraph{Two species predator prey model.} 
The model used in Figure \ref{fig:hilbert_continuation} highlighting a continuation over a Hilbert curve comes from Ref.~\cite{Zeng2024} and given by
\begin{equation*}
\begin{array}{rcl}
    \dot{x} & = & x(1-x)(x - E) - s y \\
    \dot{y} &=& y(sC - D) \\
    \mbox{where\quad} s & = & x^2/(Ax^2 + Bx + 1) 
\end{array}
\end{equation*}
with $A = 2.0551, B = -2.6, D = 1$ and $C, E$ as in Figure 1.

\paragraph{Chaotic Lorenz-84.} 
The Lorenz-84 system used in \S\ref{ex:lorenz84} is given by
\begin{align*}
\dot{x} &= -y^2 - z^2 - ax + aF, \\
\dot{y} &= xy - bxz - y + G, \label{eq: Lorenz84}\\
\dot{z} &= bxy + xz - z.
\end{align*}
with $F=6.886,\,G=1.355,\,a=0.255,\,b=4.0$ and $G$ as in Fig.~\ref{fig:lorenz84}.

\paragraph{Taxonomy model.}
In the model of Ref.~\cite{Aguade-Gorgorio2024} the abundances $x_i$ of $i \in {1, \dots, D}$ species is given by 
\begin{equation}
    \frac{\mathrm{d}x_i}{\mathrm{d}t} =  x_i \left[ \sum_j a_{ij} \frac{x_j}{x_j + g_i} - d_i -\sum_j b_{ij}x_j  \right].
    \label{eq:taxonomy}
\end{equation}
In the original paper, all parameters $a, d, b, g$ where sampled randomly from prescribed distributions.
Here we avoided this stochasticity, and made all parameters be increasing functions of $i$ with the form $v_i = (1 + (i - D/2)/D)\cdot \bar{v}$ with $\bar{v}$ the parameter value. We used $\bar{g}=1, \bar{b}=0.1$ and $\bar{a},\bar{d}$ are varying as in Fig.~\ref{fig:taxonomy}. $\bar{a},\bar{d}$  set the average growth and decay rate of species.
Off-diagonal entries for $a, b$ were sampled once from a random distribution as in Ref.~\cite{Aguade-Gorgorio2024} and set fixed at those values regardless of simulation.

\paragraph{Cloud model.}
The model used in Fig.~\ref{fig:cloud_model} is too extensive to state here.
The model setup is identical to the model used in the bottom-right panel of Figure 7 of Ref.~\cite{Datseris2026Clouds}.
Details of the model and parameter values can be found therein. 

\paragraph{Vibro-impact system.}
The model used in Fig.~\ref{fig:vibroimpact} is too extensive to state here, primarily due to all the events (discontinuities/callbacks) that are needed to implement it properly.
The model setup is identical to the model used in Figure 5 panel b of Ref.~\cite{Liu2020}.
Details of the model and parameter values can be found therein, or in our provided open code \ref{met:open}. 

% \paragraph{Stommel model.}
% The model used in Fig.~\ref{fig:rate_tipping} is given by
%\begin{align}
%    dT/dt &= \eta - T - q*T \\
%    dS/dt &= \eta_2 - \eta_3*S - q*S\\
%    q &= |T-S|
%\end{align}
%and $\eta_2 = 1, \eta_3 = 0.3$, and $\eta$ varying with time as in Fig.~\ref{fig:rate_tipping} with starting value $\eta_0 = 2.6$.

\bibliographystyle{ieeetr}
\bibliography{REFERENCES}

\end{document}